\documentclass[useAMS,usenatbib]{mn2e}
\onecolumn
\usepackage{bm}
\usepackage{amsfonts}
\usepackage{amsmath}
\usepackage{graphicx}
\usepackage{subfigure}
\usepackage{times}

\def\gsim{\;\lower4pt\hbox{${\buildrel\displaystyle >\over\sim}$}\;}
\def\lsim{\;\lower4pt\hbox{${\buildrel\displaystyle <\over\sim}$}\;}
\def\grls{\;\lower4pt\hbox{${\buildrel\displaystyle >\over <}$}\;}
\title[Dark Matter Halo and Isopedic Disc Magnetic Fields]
  {Global non-axisymmetric perturbation configurations in a composite
 disc system with an isopedic magnetic field: \newline
 relation between dark matter halo and magnetic field}
\author[M. Xiang-Gruess, Y.-Q. Lou, W. J. Duschl]
  {Meng Xiang-Gruess$^{1}$, Yu-Qing Lou$^{2,\ 3}$ and Wolfgang J. Duschl$^{1,\ 4}$\\
$^1$ Christian-Albrechts-Universit\"at zu Kiel, Institut f\"ur Theoretische Physik
     und Astrophysik, Leibnizstr. 15, 24118 Kiel, Germany\\
$^2$ Physics Department and Tsinghua Centre for Astrophysics
(THCA), Tsinghua University, Beijing, 100084, China\\
$^3$ National Astronomical Observatories, Chinese Academy of Sciences,
     A20, Datun Road, Beijing 100012, China\\
$^4$ Steward Observatory, The University of Arizona, 933 North Cherry Ave,
     Tucson, AZ 85721, USA}
\date{Accepted ...
    Received ....}
\pagerange{\pageref{firstpage}--\pageref{lastpage}}
\pubyear{2009}

\def\LaTeX{L\kern-.36em\raise.3ex\hbox{a}\kern-.15em
    T\kern-.1667em\lower.7ex\hbox{E}\kern-.125emX}

\begin{document}
\label{firstpage}
\maketitle

\begin{abstract}
We study global non-axisymmetric stationary perturbations of aligned
and unaligned logarithmic spiral configurations in an axisymmetric
composite differentially rotating disc system of scale-free stellar
and isopedically magnetized gas discs coupled by gravity. The
infinitely thin gas disc is threaded across by a vertical magnetic
field $B_z$ with a constant dimensionless isopedic ratio
$\lambda\equiv 2\pi\sqrt{G} \Sigma^{(g)}/B_z$ of surface gas mass
density $\Sigma^{(g)}$ to $B_z$ with $G$ being the gravitational
constant. Our exploration focuses on the relation between the
isopedic ratio $\lambda$ and the dark matter amount represented by a
gravitational potential ratio $f\equiv\overline{\Phi}/\Phi$ in order
to sustain stationary perturbation configurations, where
$\overline{\Phi}$ is the gravitational potential of a presumed
axisymmetric halo of dark matter and $\Phi$ is the gravitational
potential of the composite disc matter. For typical disc galaxies,
we explore relevant parameter ranges numerically. High and low
$\lambda$ values correspond to relatively weak and strong magnetic
fields given the same gas surface mass density, respectively. The
main goal of our model analysis is to reveal the relation between
isopedic magnetic fields and dark matter halo in spiral galaxies
with globally stationary perturbation configurations. Our results
show that for stationary perturbation configurations, fairly strong
yet realistic magnetic fields require a considerably larger amount
of dark matter in aligned and unaligned cases than weak or moderate
magnetic field strengths. We discuss astrophysical and cosmological
implications of our findings. For examples, patterns and pattern
speeds of galaxies may change during the course of galactic
evolution. Multiple-armed galaxies may be more numerous in the early
Universe. Flocculent galaxies may represent the transitional phase
of pattern variations in galaxies.
\end{abstract}

\begin{keywords}
galaxies: haloes ---  galaxies: ISM --- galaxies: kinematics and
dynamics --- magnetic fields --- magnetohydrodynamics (MHD) ---
waves
\end{keywords}

\section{Introduction}

Since the seminal studies of the galactic density wave theory in the
1960's pioneered by Lin \& Shu (1964, 1966), many advanced works
have been done to describe the large-scale dynamics of
`grand-design' spiral structures in disc galaxies. In terms of
theoretical development, it is natural to first concentrate
on linear perturbation theories. As synchrotron radio observations
have revealed since the early 1970's
that magnetic fields are ubiquitous in disc galaxies (e.g.
Sofue, Fujimoto \& Wielebinski 1986; Beck et al. 1996; 
Zweibel \& Heiles 1997; Brown, Taylor, Wielebinski \& Mueller 2003),
magnetic fields have been recognized as an important aspect in the
model treatment of spiral galaxies and thus the magnetohydrodynamic
(MHD) density wave theory was developed (Fan \& Lou 1996; Lou \& Fan
1998). Over the past decade, a research area has been developed on
the so-called scale-free discs and perturbation structures therein
(e.g. Syer \& Tremaine 1996; Shu et al. 2000; Shen \& Lou 2004; Lou
\& Zou 2004; Shen, Liu \& Lou 2005; Lou \& Zou 2006; Wu \& Lou 2006;
Lou \& Bai 2006). The scale-free condition corresponds to a
power-law dependence of all physical quantities on the cylindrical
radius $r$, for example, the background surface mass density is
$\Sigma_0\propto r^{-2\beta-1}$ with $\beta$ being a constant
scaling index. Such analysis was first initiated for relatively
simple models of only one gas disc. We are now able to explore a
composite system of a stellar disc and an isopedically magnetized
gas disc embedded in a dark matter halo. For this overall
configuration, we study possible large-scale perturbation structures
and stationary MHD density waves. To simplify the mathematical
treatment, the stellar disc is approximated as a fluid and the
magnetized gaseous disc as a magnetofluid and both are assumed to be
geometrically razor-thin and scale-free discs. In general
situations, the scaling exponent $\beta$ may vary radially in a
certain range for different disc galaxies. For more or less flat
rotation curves which can be grossly determined in many spiral
galaxies from large radii to the centre (e.g. Rubin 1965; Roberts \&
Rots 1973; Krumm \& Salpeter 1976, 1977; Bosma 1981; Kent 1986,
1987, 1988), index $\beta$ may be regarded as almost zero.
According to the classical Newtonian gravity theory, a significantly
larger amount of mass than the visible mass of the disc is needed in
order to sustain observed grossly flat rotation curves. As a
consequence of this discrepancy between the visible and the required
mass as well as the disc stability property, the presence of a
massive dark matter halo has been proposed (e.g. Ostriker \& Peebles
1973; Ostriker et al. 1974; Binney \& Tremaine 1987). Nowadays, the
inclusion of a massive axisymmetric dark matter halo is a common
practice for studying large-scale galactic dynamics.

Meanwhile, observations of disc galaxies nowadays can be used to
estimate open magnetic field strengths and configurations on both
sides of a galactic disc plane. While these open magnetic fields are
most certainly interlaced with closed coplanar magnetic fields, at
this stage of our theoretical model development, we consider only
open magnetic field crossing the disc plane almost
vertically.\footnote{The studies of coplanar magnetic fields in a
rotating thin disc system can be found in Lou \& Zou (2004, 2006)
and Lou \& Bai (2006).} Shu \& Li (1997) introduced the so-called
isopedic condition of $B_z(r,\ \theta,\ t)\propto\Sigma^{(g)}$ where
$B_z$ is the $z-$component of the magnetic field and $\Sigma^{(g)}$
is the gas disc surface mass density.
The ratio
$\Sigma^{(g)}/B_z$ thus remains constant in time and space. This
situation is referred to as the isopedic magnetic field
configuration. In fact, this constant ratio can be actually inferred
by the magnetic field-freezing effect according to the ideal
magnetohydrodynamic (MHD) equations (Lou \& Wu 2005).
Effective
modifications to the gravitational potential and the gas pressure
are derived in Shu \& Li (1997) for a single isopedically magnetized
gas disc. This isopedic condition was first proposed as an {\it
ansatz}
based on numerical simulations of magnetized cloud core formation
through ambipolar diffusions (e.g. Nakano 1979; Lizano \& Shu 1989;
and references cited in Shu \& Li 1997).
On the basis of earlier work of Shu
\& Li (1997), Lou \& Wu (2005) generalize their results into
theorems for two gravitationally coupled discs where only the gas
disc is isopedically magnetized. Moreover, they actually demonstrate
that a constant ratio $\lambda=2\pi\sqrt{G}\Sigma^{(g)}/B_z$, with
$G$ being the gravitational constant, can be proved as a consequence
of frozen-in condition by the standard ideal nonlinear MHD
equations.
In this model analysis, we shall adopt the same expressions of Lou
\& Wu (2005) for the effect of an isopedic magnetic field anchored
in the gas disc component.

In order to catch the essence of MHD density wave equations from
analytical and numerical explorations, a few assumptions and
simplifications are invoked such as the stationary condition in
our frame of reference. As first advanced and hypothesized by Lin
\& Shu (1966), the quasi-stationary spiral structure (QSSS) has
been studied since the early 1960's (e.g. Bertin \& Lin 1996 and
extensive references therein). There the pattern speed of density
wave travels very slowly as compared to the differential rotation
speed $v_\theta$ of the disc matter. Our stationary requirement
can be simply regarded as a limiting case of the QSSS
approximation. Based on observations of the nearby spiral galaxy
M81, Visser (1980a, b) has shown that the density wave theory does
indeed account for gross aspects of the large-scale dynamics of
this spiral galaxy.

Kendall et al. (2008) presented further observational data
analysis of the nearby galaxy M81, where the large-scale spatial
offset between the (old) stellar spiral wave and the gas shock
front was examined. Theoretical considerations (e.g. Gittins \&
Clarke 2004; Chakrabarti 2008) have shown that the offset between
the gas shock front (i.e. the maximum gas density) and the spiral
potential minimum (determined by the spiral pattern in the old
stellar disc component) may yield clues about the lifetime of a
spiral pattern. By comparing observational results with the
predictions of Gittins \& Clarke (2004), Kendall et al. (2008)
found that the galaxy M81 probably possesses a long-lived spiral
pattern. Very recently, Chakrabarti (2008) performed numerical
simulations to demonstrate that an angular momentum transport
takes place from gas to stars in a composite system, leading to an
approximately time-steady spiral structure in the stellar disc
component (treated as collisionless N-body particles). A major
part that plays an important role is the relatively cold gas disc
particles modelled by smoothed particle hydrodynamic (SPH)
and is dissipative due to an artificial viscosity. Regarding the
spiral structure in the gaseous disc component, Chakrabarti et al.
(2003) have shown, that the so-called ultraharmonic resonances that
the gas experiences in a fixed stellar spiral potential can lead to
highly time-dependent gas responses, not allowing long-lived spiral
patterns in the gas disc. In our model of a composite disc system,
there is no fixed stellar potential but rather stellar and gaseous
discs are gravitationally coupled to each other in terms of
dynamical adjustment. While the simulation results reveal only a
long-lived stellar spiral pattern, for our semi-analytical
calculations, a stationary requirement for both spiral patterns in
stellar and gaseous discs may be imposed without inconsistency. This
offers a useful construct for modelling large-scale patterns in
isopedically magnetized spiral galaxies.

Theoretically, two classes of stationary MHD density wave
solutions with in-phase and out-of-phase density perturbations are
expected to exist in a composite disc system according to Lou \&
Shen (2003).
The physical reason for these two different couplings between the
stellar and gaseous discs is a dynamic interplay between star
formation and gas clumping processes (Shen \& Lou 2004). When the
surface mass densities are out of phase, the self-gravity of the
composite disc system is reduced. The azimuthal propagation of
density wave is therefore faster and if a stationary configuration
is desired, then the disc should rotate faster. When the surface
mass densities are in phase, the self-gravity is enhanced. The
azimuthal propagation of the density wave is slower and the disc
should then rotate slower in order to maintain a stationary
configuration (Lou \& Shen 2003).

Our present investigation mainly focuses on the variation of the
amount of dark matter in the halo needed to sustain stationary
perturbation configurations as the isopedic magnetic field
strength varies. Therefore, we use typical values of disc galaxies
for different parameters such as the `sound speed' or the surface
mass density.
By invoking several physical requirements for the dark matter halo
such as a non-negative potential ratio and so forth, the physical
parameter range of stationary perturbation solutions can be
identified.

In Section \ref{sec:nomenclature}, the problem formulation, the
basic formulae and the coupled fluid and magnetofluid equations for
the two discs are presented. In Section \ref{sec:magnetic field}, we
discuss effective modifications to the gravitational potential and
the enthalpy in the gas disc when an isopedic magnetic field is
introduced. In Section \ref{sec:equilibrium}, we derive several
relevant relations for the background rotational equilibrium
configuration. In Section \ref{sec:MHD perturbations}, MHD
perturbations are introduced and the relevant global stationary
dispersion relation is derived. For globally stationary MHD
perturbation configurations, we
calculate the dark matter amount and discuss which parameters affect
this amount. Section \ref{sec:parameters} describes briefly the
values for the parameters adopted for this model analysis. In
Section \ref{sec:nu=0}, we study aligned cases where perturbations
of different radii can be aligned radially. In Section
\ref{sec:nu!=0}, we study unaligned logarithmic spiral cases with a
constant radial flux of angular momentum. Such unaligned cases have
MHD perturbations which are systematically phase shifted azimuthally
as a function of radius $r$. In unaligned cases, MHD density waves
propagate in both azimuthal and radial directions. The constant
radial flux of angular momentum is realized by setting the
perturbation scale-free index $\beta_1$ to $1/4$ (e.g. Goldreich \&
Tremaine 1979; Shen et al. 2005). In Section \ref{sec:limits}, we
discuss two limiting cases of no magnetic field and only a single
isopedically magnetized gaseous disc. In the last Section 10, we
summarize our work and draw conclusions.

\section{Nomenclature and Basic Fluid and Magnetohydrodynamic (MHD) Equations}
\label{sec:nomenclature}

We first introduce notations for the relevant physical variables
used in our model formulation and analysis. We largely adopt the
same nomenclature of Wu \& Lou (2005) for a model of a single
scale-free disc with an isopedic magnetic field. Whenever the
superscript $(i)$ is attached to a physical variable, the relevant
equation is valid for both stellar $(s)$ and gaseous $(g)$ discs in
parallel. Cylindrical coordinates $(r,\ \theta,\ z)$ are adopted for
our model formulation and we confine our analysis within the $z=0$
plane for an infinitely thin composite disc system in differential
rotation. For example, variable $\Sigma^{(s)}(r,\ \theta,\ t)$ is
the two-dimensional surface mass density of the stellar disc and
$\Sigma^{(g)}(r,\ \theta,\ t)$ is that of the isopedically
magnetized gas disc. The total surface mass density $\Sigma(r,\
\theta,\ t)=\Sigma^{(s)}+\Sigma^{(g)}$ is the sum of these two disc
surface mass densities. Two different coplanar bulk flow velocities
are possible in the two discs: the radial bulk flow velocity
$v^{(i)}_r(r,\ \theta,\ t)$ and the azimuthal bulk flow velocity
$v^{(i)}_\theta(r,\ \theta,\ t)$. For the background rotational
equilibrium disc configuration of axisymmetry, the radial flow
velocities should vanish. The two-dimensional pressure (i.e.
vertically integrated pressure) in each disc is described by
$\Pi^{(i)}(r,\ \theta,\ t)$. Between the two-dimensional variables
and their three-dimensional counterparts, the following integral
relations exist,
\begin{eqnarray}
\Pi^{(i)}(r,\ \theta,\ t)=\int\limits_{\Delta z}p^{(i)}(r,\ \theta,\
z,\ t)dz\hspace{1 cm}
\hbox{ and }\hspace{1 cm}\Sigma^{(i)}(r,\ \theta,\
t)=\int\limits_{\Delta z}\rho^{(i)}(r,\ \theta,\ z,\ t)dz\ ,
\end{eqnarray}
where $p^{(i)}$ is the three-dimensional pressure and $\rho^{(i)}$
is the three-dimensional mass density, and $\Delta z$ is the disc
thickness presumed to be very small compared to radius $r$. Thus,
$\Pi^{(i)}$ is the vertical integral of pressure $p^{(i)}$. The
three-dimensional polytropic equation of state is
$p^{(i)}=k^{(i)}_3[\rho^{(i)}]^\gamma$
with $\gamma$ being the polytropic index and $k^{(i)}_3$ being
related to the disc entropy. The vertically integrated barotropic
equation of state can then be expressed as
\begin{equation}
\Pi^{(i)}=k^{(i)}\left[\Sigma^{(i)}\right]^n\ ,\label{eq:Pi^i}
\end{equation}
where index $n>0$ is the barotropic index, $k^{(i)}>0$ for a {\it
warm} disc and $k^{(i)}=0$ for a {\it cold} disc.
The barotropic sound speed $a^{(i)}$ in each disc is then defined
by
\begin{equation}
\left[a^{(i)}\right]^2={\rmn{d}\Pi^{(i)}}\big /{\rmn{d}
\Sigma^{(i)}}=nk^{(i)}\left[\Sigma^{(i)}\right]^{n-1}
=n\Pi^{(i)}\big/\Sigma^{(i)}\ .\label{eq:a^2}
\end{equation}
The `sound speed' $a^{(s)}$ in the stellar disc is effectively
related to the velocity dispersion of stars. The enthalpy $H^{(i)}$
in each disc is given by
\begin{equation}
H^{(i)}=\int\frac{\rmn{d}\Pi^{(i)}}{\Sigma^{(i)}}
=\frac{nk^{(i)}[\Sigma^{(i)}]^{n-1}}{(n-1)}
=\frac{[a^{(i)}]^2}{(n-1)}\ \label{eq:enthalpy}
\end{equation}
and is proportional to the square of `sound speed' in each disc.
The gravitational coupling between the stellar and magnetized
gaseous discs is described by the well-known Poisson integral
\begin{equation}
\Phi(r,\theta, t)=\oint\rmn{d}\psi\int\frac{-G\Sigma(r',\ \psi,\
t)r'\rmn{d}r'}{[r'^2+ r^2-2rr'\cos(\psi-\theta)]^{1/2}}\ ,
\label{eq:Poisson}
\end{equation}
where $G=6.67\times 10^{-8}\hbox{ g}^{-1}\hbox{ cm}^{3}\hbox{
s}^{-2}$ is the gravitational constant. The gravitational potentials
of the two discs can be separately written as
\begin{eqnarray}
\Phi^{(s)}(r, \theta, t)=\oint\rmn{d}\psi\int
\frac{-G\Sigma^{(s)}(r',\ \psi,\ t)r'\rmn{d}r'}{[r'^2+
r^2-2rr'\cos(\psi-\theta)]^{1/2}}\ ,\qquad\qquad
\Phi^{(g)}(r,\theta, t)=\oint\rmn{d}\psi\int
\frac{-G\Sigma^{(g)}(r',\ \psi,\ t)r'\rmn{d}r'}{[r'^2+
r^2-2rr'\cos(\psi-\theta)]^{1/2}}\ ,
\end{eqnarray}
with $\Phi=\Phi^{(s)}+\Phi^{(g)}$. In a disc galaxy, an
axisymmetric massive dark matter halo is represented by a
gravitational potential $\overline{\Phi}$.
A recent numerical exploration (e.g. Diemand et al. 2008) has
shown that the dark matter halo is not smooth and uniform, but
rather consists of many smaller clumpy subhalos. For simplicity,
we assume the dark matter halo to be grossly axisymmetric on large
scales and unperturbed in our model analysis at this stage.
The composite system of two coupled discs with an axisymmetric
dark matter halo is described by coupled nonlinear fluid equations
$(\ref{eq:conti})-(\ref{eq:euler2})$ below.
\begin{eqnarray}
\frac{\partial\Sigma^{(i)}}{\partial t}+\frac{1}{r}
\frac{\partial}{\partial r}\left[r\Sigma^{(i)}v^{(i)}_r\right]+
\frac{1}{r}\frac{\partial}{\partial\theta}
\left[\Sigma^{(i)} v^{(i)}_\theta\right]=0\ ,\qquad\qquad \label{eq:conti}\\
\frac{\partial v^{(i)}_r}{\partial t}+v^{(i)}_r\frac{\partial
v^{(i)}_r}{\partial r}+\frac{v^{(i)}_\theta}{r}\frac{\partial
v^{(i)}_r} {\partial \theta}-\frac{\big[v^{(i)}_\theta\big]^2}{r}
=-\frac{1}{\Sigma^{(i)}}\frac{\partial \Pi^{(i)}} {\partial
r}-\frac{\partial (\Phi+\overline{\Phi})}{\partial r}
=-\frac{\partial\big[H^{(i)}+\Phi+\overline{\Phi}\big]}
{\partial r}\ ,\qquad\qquad\label{eq:euler1}\\
\frac{\partial v^{(i)}_\theta}{\partial t}
+v^{(i)}_r\frac{\partial v^{(i)}_\theta}{\partial r}+
\frac{v^{(i)}_\theta}{r}\frac{\partial v^{(i)}_\theta}
{\partial\theta}+\frac{v^{(i)}_\theta v^{(i)}_r}{r}
=-\frac{1}{\Sigma^{(i)}}\frac{\partial\Pi^{(i)}}{r\partial
\theta}-\frac{\partial (\Phi+\overline{\Phi})}{r\partial\theta}
= -\frac{\partial \big[H^{(i)}+\Phi+\overline{\Phi}\big]}{r
\partial \theta}\ \qquad\qquad\label{eq:euler2}
\end{eqnarray}
(e.g. Shen \& Lou 2004). Equation (\ref{eq:conti}) is the mass
conservation, and equations (\ref{eq:euler1}) and
(\ref{eq:euler2}) are the radial and azimuthal momentum equations.
These fluid equations for both stellar and gaseous discs are
confined to the two-dimensional disc plane at $z=0$, while Poisson
integral equation (\ref{eq:Poisson}) is three-dimensional.
These equations do not involve an isopedic magnetic field at this
stage.

\subsection{Scale-Free Discs and Relevant Parameter Ranges}

Scale-free discs are represented by power-law forms with a few
disc index parameters. We write a self-consistent form of disc
solution as
\begin{eqnarray}
v_{r}^{(g)}=e^{(g)}(\varphi)r^{-\beta}\ ,\qquad\qquad
v_{r}^{(s)}=e^{(s)}(\varphi)r^{-\beta}\ ,\\
v_{\theta}^{(g)}=b^{(g)}(\varphi)r^{-\beta}\ ,\qquad\qquad
 v_{\theta}^{(s)}=b^{(s)}(\varphi)r^{-\beta}\ ,\\
\Sigma^{(g)}=S^{(g)}(\varphi)r^{-2 \beta-1}\ ,\qquad\qquad
\Sigma^{(s)}=S^{(s)}(\varphi)r^{-2\beta-1}\ ,\label{eq:sigmas_0} \\
\Pi^{(g)}=k^{(g)}\big[\Sigma^{(g)}\big]^n\ ,\qquad\qquad
\Pi^{(s)}=k^{(s)}\big[\Sigma^{(s)}\big]^n\ ,\\
H^{(g)}=Q^{(g)}(\varphi)r^{-2\beta}\ ,\qquad\qquad
H^{(s)}=Q^{(s)}(\varphi)r^{-2\beta}\,\label{eq:H^{(g)},H^{(s)}} \\
\Phi=-P(\varphi) r^{-2\beta}\ ,\qquad\qquad\qquad
\overline{\Phi}=-\overline{P}r^{-2\beta}\ ,
\end{eqnarray}
where $e^{(i)}$,
$b^{(i)}$, $S^{(i)}$, $Q^{(i)}$, and $P$ are functions of only
$\varphi$ which is an argument abbreviation
$\varphi\equiv\theta+\mu \ln r$ with $\mu$ being a parameter, and
$\overline P$ for the dark matter halo is a constant coefficient
independent of $\varphi$. The detailed mathematical procedure of
constructing such self-similar scale-free solutions can be found
in Lynden-Bell \& Lemos (1999). By comparing equations
(\ref{eq:enthalpy}) and (\ref{eq:H^{(g)},H^{(s)}}), we immediately
obtain a relation between the two power-law indices $n$ and
$\beta$, namely
\begin{equation}
(2\beta +1)(n-1)=2\beta\ ,\hspace{0.5cm}\to\hspace{0.5cm}
n={(1+4\beta)}/{(1+2\beta)}\ .
\end{equation}
Based on expressions (\ref{eq:Pi^i}) and (\ref{eq:a^2}), we find
that for warm discs with $k^{(i)}>0$, the barotropic index $n$ must
also be positive in order to ensure a positive right-hand side (RHS)
of equation (\ref{eq:a^2}) (i.e. a real sound speed). This physical
requirement for warm discs leads to either $\beta>-1/4$ or
$\beta<-1/2$. For cold discs with $k^{(i)}\to 0$, there is no
constraint on the barotropic index $n$. Another empirical constraint
which arises from observational results is that $\Sigma^{(i)}$
should decrease with increasing $r$ which requires a $\beta$ larger
than $-1/2$. With the assumption of decreasing $\Sigma^{(i)}$, a
singularity of mass would arise in the central region as
$r\rightarrow 0^{+}$. This problem can be remedied by requiring a
finite integral
\begin{eqnarray}\label{eq:claim}
\lim_{r\to 0}\int_0 ^r\int_0 ^{2\pi}\Sigma^{(i)}(r,\theta) r
\rmn{d}r\rmn{d}\theta
=\int_0^{2\pi}S^{(i)}(\theta)\rmn{d}\theta \lim_{r\to
0}\left[\frac{r^{1-2\beta}}{1-2\beta}\right]_0^r < +\infty\
\end{eqnarray}
and this then leads to $\beta<1/2$. Therefore for warm discs with
$k^{(i)}>0$, the scaling index range is $\beta\in (-1/4,\ 1/2)$
and for cold discs with $k^{(i)}=0$, we simply require $\beta\in
(-1/2,\ 1/2)$.


\section{An Isopedic magnetic field across a scale-free
thin gaseous disc}\label{sec:magnetic field}

Effective modifications of the gravitational potential and the
pressure in a composite system of two discs due to the very
presence of an isopedic magnetic field were derived in reference
to the study of singular isothermal discs (SIDs) by Lou \& Wu
(2005). SIDs represent only a special class (i.e. $\beta=0$) of
scale-free discs in our more general formalism here. We now
briefly summarize their results and introduce the relevant
variables in the current model context. Following Li \& Shu
(1996), we define a constant dimensionless ratio $\lambda$ as
\begin{equation}
\lambda=2\pi G^{1/2}\Sigma^{(g)}/B_z\equiv 2\pi G^{1/2}
\Lambda=\rmn{constant}>0\ .\label{eq:lambda1}
\end{equation}
Parameters $\lambda$ and $\Lambda$ indicate how strong the
magnetic field is as compared to the surface mass density of the
gas disc; they are proportional to $B_z^{-1}$ such that for weak
or strong magnetic fields they become very large or small,
respectively.
Parameter $\eta$ is a dimensionless ratio of the total horizontal
gravitational acceleration
$|\vec{f}_\parallel|=|\vec{f}^{(g)}_\parallel+\vec{f}^{(s)}_\parallel|$
(continuous across the thin disc along the vertical direction) to
the total vertical gravitational acceleration just above the two
discs with a total surface mass density
$\Sigma=\Sigma^{(g)}+\Sigma^{(s)}$, namely
\begin{equation}
\eta\equiv\frac{|\vec{f}_\parallel|}{2\pi G\Sigma}
=\frac{|\vec{f}^{(g)}_\parallel|}{2\pi G\Sigma^{(g)}}\
,\label{eq:eta1}
\end{equation}
where the gaseous disc contribution to $\vec{f}_\parallel$ is
$\vec{f}^{(g)}_\parallel=-\vec{\nabla}_\parallel\Phi^{(g)}$. The
two-dimensional gradient operation $\vec{\nabla}_\parallel$ within
the disc plane coincident with $z=0$ and in terms of cylindrical
coordinates $(r,\ \theta,\ z)$ is simply
\begin{equation}
\vec{\nabla}_\parallel=\frac{\partial}{\partial r}\vec{e}_r+
\frac{1}{r}\frac{\partial}{\partial\theta}\vec{e}_\theta\ ,
\end{equation}
where $\vec{e}_r$ and $\vec{e}_\theta$ are unit vectors along the
radial and azimuthal directions, respectively. The sum of the
magnetic tension force and the horizontal gravity force acting on
the isopedically magnetized gaseous disc is
\begin{equation}
\vec{f}^{(g)}=\vec{f}^{(g)}_{\rmn{ten}}
+\vec{f}^{(g)}_\parallel=\epsilon \vec{f}^{(g)}_\parallel\
,\label{eq:force}
\end{equation}
with $\vec{f}^{(g)}_{\rmn{ten}}$ as the magnetic tension force
acting in the magnetized gaseous disc and $\epsilon\equiv
1-1/\lambda^2$. In other words, the magnetic tension force
$\vec{f}^{(g)}_{\rmn{ten}}$ and the horizontal gravity of gaseous
disc $\vec{f}^{(g)}_\parallel$ act always in opposite directions.
Effectively, equation (\ref{eq:force}) leads to the modification
of the gravitational potential for the gaseous disc as
\begin{equation}
\Phi^{(g)}\to\epsilon\Phi^{(g)}\ ,\label{epsilon}
\end{equation}
where $\epsilon$ is referred to as the {\it reduction factor}
because it is always less than unity and may become negative. The
situation of $\epsilon<0$ happens when the magnetic tension force
overwhelms the gas disc horizontal gravity. Due to the additional
magnetic pressure, the total pressure in the magnetized gaseous
disc is effectively enhanced by
\begin{eqnarray}
\Pi^{(g)}\to \Theta \Pi^{(g)}\ ,\hspace{1.0
cm}\rmn{where}\hspace{1.0
cm}\Theta=1+\frac{1+\eta^2}{\hat{\lambda}^2+\eta^2}>0\
,\hspace{1.0 cm} \hat{\lambda}=\big(1+\delta^{-1}\big)\lambda\
,\hspace{1.0 cm} \delta=\Sigma^{(g)}/\Sigma^{(s)}\ .\label{Btheta}
\end{eqnarray}
Here, $\Theta$ is referred to as the {\it enhancement factor}
because we always have $\Theta>1$. For the effective modification
of disc enthalpy $H^{(g)}$, we assume for simplicity that $\Theta$
is independent of the cylindrical coordinates $r$ and $\theta$ and
then get $H^{(g)}\to\Theta H^{(g)}$.

In summary, the gravitational potential $\Phi^{(g)}$, the pressure
$\Pi^{(g)}$ and the enthalpy $H^{(g)}$ in an isopedically
magnetized gaseous disc are effectively modified according to
$\Phi^{(g)}\to\epsilon \Phi^{(g)}$, $\Pi^{(g)}\to\Theta \Pi^{(g)}$
and $H^{(g)}\to\Theta H^{(g)}$ while the counterparts of these
variables remain unchanged in the stellar disc. The magnetosonic speed squared
in the isopedically magnetized gaseous disc is then
\begin{equation}
\Theta\frac{\rmn{d}\Pi^{(g)}}{\rmn{d}\Sigma^{(g)}} =n\Theta
k^{(g)}\left[\Sigma^{(g)}\right]^{n-1}=\Theta\left[a^{(g)}\right]^2\
.
\end{equation}
In reference to equations
$(\ref{eq:conti})-(\ref{eq:euler2})$, an isopedic
magnetic field in the gaseous disc leads to the following modified
set of coupled equations.
\begin{eqnarray}
\frac{\partial\Sigma^{(i)}}{\partial t}+\frac{1}{r}
\frac{\partial}{\partial r}\left[r\Sigma^{(i)} v^{(i)}_r\right]+
\frac{1}{r}\frac{\partial}{\partial\theta}\left[\Sigma^{(i)}
v^{(i)}_\theta\right]=0 \ ,\label{eq27}\\
\frac{\partial v^{(s)}_r}{\partial t}+v^{(s)}_r \frac{\partial
v^{(s)}_r}{\partial r}+\frac{v^{(s)}_\theta}{r} \frac{\partial
v^{(s)}_r}{\partial\theta}-\frac{\big[v^{(s)}_\theta\big]^2}{r}
=-\frac{1}{\Sigma^{(s)}}\frac{\partial \Pi^{(s)}}{\partial
r}-\frac{\partial
\big[\Phi^{(s)}+\Phi^{(g)}+\overline{\Phi}\big]}{\partial
r}
=-\frac{\partial
\big[H^{(s)}+\Phi^{(s)}+\Phi^{(g)}
+\overline{\Phi}\big]}{\partial r}\ ,\label{eq:Euler1s}\\
\frac{\partial v^{(g)}_r}{\partial t}+v^{(g)}_r \frac{\partial
v^{(g)}_r}{\partial r}+ \frac{v^{(g)}_\theta}{r} \frac{\partial
v^{(g)}_r} {\partial \theta}-\frac{\big[v^{(g)}_\theta\big]^2}{r}
= -\frac{1}{\Sigma^{(g)}}\frac{\partial \big[\Theta
\Pi^{(g)}\big]}{\partial r}- \frac{\partial
\big[\Phi^{(s)}+\epsilon
\Phi^{(g)}+\overline{\Phi}\big]}{\partial r}
=-\frac{\partial \big[\Theta H^{(g)}+\Phi^{(s)}+\epsilon
\Phi^{(g)}+\overline{\Phi}\big]}{\partial r}\ ,
\label{eq:Euler1g}\\
\frac{\partial v^{(s)}_\theta}{\partial t}+v^{(s)}_r
\frac{\partial v^{(s)}_\theta}{\partial r}+
\frac{v^{(s)}_\theta}{r}\frac{\partial v^{(s)}_\theta}
{\partial\theta}+\frac{v^{(s)}_\theta v^{(s)}_r}{r}
=-\frac{1}{\Sigma^{(s)} r}\frac{\partial
\Pi^{(s)}}{\partial\theta}-\frac{\partial
\big[\Phi^{(s)}+\Phi^{(g)}+\overline{\Phi}\big]}{r
\partial \theta}
=-\frac{\partial \big[H^{(s)}+\Phi^{(s)}+\Phi^{(g)}
+\overline{\Phi}\big]}{r \partial\theta}\ ,\\
\frac{\partial v^{(g)}_\theta}{\partial t}+v^{(g)}_r\frac{\partial
v^{(g)}_\theta}{\partial r}+\frac{v^{(g)}_\theta}{r}\frac{\partial
v^{(g)}_\theta}{\partial\theta}+\frac{v^{(g)}_\theta
v^{(g)}_r}{r}
= -\frac{1}{\Sigma^{(g)} r}\frac{\partial \big[\Theta
\Pi^{(g)}\big]}{\partial\theta}-\frac{\partial
\big[\Phi^{(s)}+\epsilon\Phi^{(g)}+\overline{\Phi}\big]}{r\partial
\theta}
= -\frac{\partial\big[\Theta H^{(g)}+\Phi^{(s)}+\epsilon
\Phi^{(g)}+\overline{\Phi}\big]}{r
\partial \theta}\ .\label{eq31}
\end{eqnarray}
Equations $(\ref{eq27})-(\ref{eq31})$ bear very similar form of
hydrodynamic equations $(\ref{eq:conti})-(\ref{eq:euler2})$ with the
effect of an isopedic magnetic field being subsumed into two
dimensionless parameters $\Theta$ and $\epsilon$ related to the
magnetized gaseous disc component (Shu \& Li 1997; Wu \& Lou 2006).

\section{Equilibrium configuration of a composite
rotating disc system}\label{sec:equilibrium}

In a stationary equilibrium, the stellar and gas discs in a
composite disc system rotate with {\it different} angular speeds in
general while satisfying the basic nonlinear fluid-magnetofluid
equations.
For both discs in coupled rotational equilibrium
of axisymmetry, the gravitational acceleration caused by the dark
matter halo and the two discs together is the same in the two radial
force balance conditions [i.e. eqns (\ref{eq:1.}) and (\ref{eq:2.})
below]. Meanwhile, the gas pressure and magnetic Lorentz forces
together in the gaseous disc are different from the effective
pressure force produced by stellar velocity dispersion in the
stellar disc in general. These naturally lead to two different
angular speeds of the two discs.
Equilibrium variables of a composite
rotating disc configuration are denoted by a subscript $0$. For a
rotating disc configuration in a stationary axisymmetric
equilibrium, radial velocities vanish with
$v^{(g)}_{r0}=v^{(s)}_{r0}=0$ and $v^{(g)}_{\theta
0}=\Omega^{(g)}_0 r$, $v^{(s)}_{\theta 0}=\Omega^{(s)}_0r$ with
$\Omega^{(i)}_0\equiv\Omega^{(i)}_0(r)$ being the differential
angular rotation speed of each disc. The mass conservation
equations are consistently satisfied by our prescription. Radial
momentum equations (\ref{eq:Euler1s}) and (\ref{eq:Euler1g}) then
lead to two balance conditions
\begin{eqnarray}
\frac{\big[v^{(s)}_{\theta 0}\big]^2}{r}=\frac{\partial
\big[H_0^{(s)}+\Phi^{(s)}_0
+\Phi^{(g)}_0+\overline{\Phi}_0\big]}{\partial r}\ ,\label{eq:1.}\\
\frac{\big[v^{(g)}_{\theta 0}\big]^2}{r}=\frac{\partial [\Theta
H_0^{(g)}+\Phi^{(s)}_0+\epsilon
\Phi^{(g)}_0+\overline{\Phi}_0]}{\partial r}\ .\label{eq:2.}
\end{eqnarray}
In general, the radial derivative of disc enthalpy is related to
the `sound speed' as follows
\begin{eqnarray}
\frac{\partial H_0^{(i)}}{\partial r}=\frac{\partial }{\partial r}
\left[\frac{n k^{(i)}}{(n-1)}\big[S_0^{(i)}\big]^{n-1}
r^{(-2\beta-1)(n-1)}\right]
= -\frac{\big[a^{(i)}\big]^2}{r}(2\beta+1)\ .
\end{eqnarray}
For the total surface mass density in a rotational equilibrium
configuration, we simply take
$\Sigma_0=\Sigma^{(g)}_0+\Sigma^{(s)}_0=S^{(g)}_0
r^{-2\beta-1}+S^{(s)}_0r^{-2\beta-1}=S_0 r^{-2\beta-1}$ where
$S^{(s)}_0$ and $S^{(g)}_0$ are two positive constant coefficients
(see also equation \ref{eq:sigmas_0}) and
$S_0=S^{(s)}_0+S^{(g)}_0$ is the sum of these two coefficients
$S^{(s)}_0$ and $S^{(g)}_0$. Prescribed as such, one can make use
of formulae in Qian (1992) for the total gravitational potential
\begin{equation}
\Phi_0=-Gr\Sigma_0 Y_0(\beta)=-GY_0(\beta)S_0r^{-2\beta}\
,\label{eq:phi(sigma)}
\end{equation}
where the coefficient factor $Y_0(\beta)$ is related to the
standard $\Gamma$ functions by
\begin{equation}\label{Y0def}
Y_0(\beta)\equiv\frac{\pi\Gamma(1/2-\beta)
\Gamma(\beta)}{\Gamma(1-\beta)\Gamma(1/2+\beta)}\ .
\end{equation}
For later global non-axisymmetric perturbation analysis, we also
introduce below a generalization of $Y_0(\beta)$ as $Y_m(\beta)$
defined by
\begin{equation}\label{Ym}
Y_m(\beta) \equiv \frac{\pi \Gamma(m/2-\beta+1/2)
\Gamma(m/2+\beta)}{\Gamma(m/2-\beta+1)\Gamma(m/2+\beta+1/2)}\ ,
\end{equation}
where $m$ is an integer in the complex phase factor
$\exp(-im\theta)$ for characterizing non-axisymmetric coplanar
perturbations. As shown by Qian (1992), the valid range for
expression (\ref{Ym}) is $-m/2<\beta<(m+1)/2$. For the smallest
$m=1$, the valid range of $\beta$ is $(-1/2,\ 1)$ which is wider
than the derived $\beta$ range $(-1/4,\ 1/2)$ for warm background
discs; and therefore we do not worry about the parameter regime of
$\beta$ in our subsequent global perturbation analysis. This leads
to the following relation for the horizontal-to-perpendicular
gravity ratio $\eta$ as defined by equation (\ref{eq:eta1}), viz.
\begin{eqnarray}
\eta=\frac{|-\vec{\nabla}_\parallel\Phi_0|}{2\pi G\Sigma_0}=
\frac{|-2G\beta Y_0(\beta)r^{-2\beta-1}S_0\vec{e}_r|}{2\pi
\rmn{G}S_0r^{-2\beta-1}}
=\frac{\beta Y_0(\beta)}{\pi}\equiv\eta(\beta)\ .
\end{eqnarray}
For each equilibrium configuration in the composite disc system,
the gravitational potential and its first radial derivative are
given by
\begin{eqnarray}
\Phi^{(i)}_0=-Gr\Sigma^{(i)}_0Y_0(\beta)
=-GY_0(\beta)r^{-2\beta}S^{(i)}_0\
\hspace{1cm}\hbox{ and }\hspace{1cm}\frac{\partial
\Phi^{(i)}_0}{\partial r}=2\beta GY_0(\beta)\Sigma^{(i)}_0\ .
\end{eqnarray}
To measure the effect of a dark matter halo, we introduce a
gravitational potential ratio parameter $f$ as
\begin{equation}
f\equiv
\frac{\overline{\Phi}_0}{\Big[\Phi^{(s)}_0+\Phi^{(g)}_0\Big]}
=\frac{\overline{\Phi}_0}{\Phi_0}\
\end{equation}
for the ratio between the dark matter halo potential to the
background potential of the two coupled discs together. The radial
derivative of the axisymmetric dark matter halo potential can now be
simply expressed as
\begin{equation}
\frac{\partial\overline{\Phi}_0}{\partial r}=f\frac{\partial
\big[\Phi^{(s)}_0+\Phi^{(g)}_0\big]}{\partial r}=f 2\beta G
Y_0(\beta)\Sigma_0\ .
\end{equation}
With the expressions for enthalpies and gravitational potentials
derived above, radial force balances (\ref{eq:1.}) and
(\ref{eq:2.}) become
\begin{eqnarray}\label{eq:1.a}
\big[v^{(s)}_{\theta 0}\big]^2+\big[a^{(s)}\big]^2(2\beta+1)
=2\beta r GY_0(\beta)
\big[\Sigma^{(s)}_0+\Sigma^{(g)}_0\big](1+f)\ ,
\\
\big[v^{(g)}_{\theta 0}\big]^2
+\Theta\big[a^{(g)}\big]^2(2\beta+1)
=2\beta rGY_0(\beta)
\left\lbrace\big[\Sigma^{(s)}_0+\Sigma^{(g)}_0\big](1+f)-(1-\epsilon)
\Sigma^{(g)}_0\right\rbrace\ .\label{eq:2.a}
\end{eqnarray}
The physical properties of the two discs are related by
equilibrium conditions (\ref{eq:1.a}) and (\ref{eq:2.a}). The two
polytropic sound speeds $a^{(s)}$ and $a^{(g)}$ in the
axisymmetric background stellar and gaseous discs are respectively
\begin{eqnarray}
\big[a^{(s)}\big]^2=nk^{(s)}\big[\Sigma^{(s)}_0\big]^{n-1}
=n\Pi^{(s)}_0\big/\Sigma_0^{(s)}
=\frac{(1+4\beta)}{(1+2\beta)} k^{(s)}
\big[S^{(s)}_0\big]^{{2\beta}/{(1+2\beta)}} r^{-2\beta}\ ,\label{eq:sss}\\
\big[a^{(g)}\big]^2=nk^{(g)}\big[\Sigma^{(g)}_0\big]^{n-1}
=n\Pi^{(g)}_0\big/\Sigma_0^{(g)}
=\frac{(1+4\beta)}{(1+2\beta)} k^{(g)}
\big[S^{(g)}_0\big]^{{2\beta}/{(1+2\beta)}}r^{-2\beta}\ .
\label{eq:ssg}
\end{eqnarray}
Combining equations (\ref{eq:1.a}) and (\ref{eq:2.a}), one can
readily deduce the equilibrium surface mass densities of two discs
as
\begin{eqnarray}
\Sigma^{(s)}_0=\frac{\big[v^{(s)}_{\theta 0}\big]^2
+\big[a^{(s)}\big]^2(2\beta+1)}{2\beta rGY_0(\beta)
(1+f) (1+\delta_0)}\ ,\qquad\ \ \label{eq:S^s}\\
\Sigma^{(g)}_0=\frac{\big[v^{(g)}_{\theta 0}\big]^2+\Theta
\big[a^{(g)}\big]^2(2\beta+1)}{2\beta rGY_0(\beta)
(f+1/\delta_0+f/\delta_0+\epsilon)}\ ,\label{eq:S^g}
\end{eqnarray}
with $\delta_0\equiv\Sigma^{(g)}_0/\Sigma^{(s)}_0$ being the ratio
of gaseous to stellar disc surface mass densities. As the surface
mass densities are positive, we obtain a necessary inequality
between the potential ratio $f$ and the ratio $\lambda$ by using
the fact that the quantities $\delta_0$, $Y_0(\beta)$ and $r$ are
all positive
\begin{eqnarray}
f+1/\delta_0+f/\delta_0+\epsilon>0\,\qquad\mathrm{for}\qquad
\beta>0\ .\label{eq:f(delta)}
\end{eqnarray}
Note that $\epsilon=1-1/\lambda^2$ may be negative for a strong
isopedic magnetic field. This is the first requirement on
potential ratio $f$ parameter which will be used later. Since
$\beta>-1/4$ for all discs, the numerators of equations
(\ref{eq:S^s}) and (\ref{eq:S^g}) are always positive which leads
to the condition that $\beta$ in the denominator must be positive
since all other factors are positive. This condition restricts our
exploration to the range of $\beta=(0,\ 1/2)$.

\section{Two-Dimensional Coplanar MHD Perturbation
Equations}\label{sec:MHD perturbations}

We now assume small coplanar perturbations with $v^{(i)}_r
=v^{(i)}_{r0}+v^{(i)}_{r 1}$, $v^{(i)}_\theta=v^{(i)}_{\theta 0}+
v^{(i)}_{\theta 1}$, $\Sigma^{(i)}=\Sigma^{(i)}_0+\Sigma^{(i)}_1$,
$\Phi^{(i)}=\Phi^{(i)}_0+\Phi^{(i)}_1$ and
$H^{(i)}=H^{(i)}_0+H^{(i)}_1$ in each disc. For these
perturbations in each disc, the following expressions of Fourier
decomposition are used
\begin{eqnarray}
v^{(i)}_{r 1}=A_r^{(i)}(r)\exp [i(\omega t-m\theta)]\ ,\\
v^{(i)}_{\theta 1}=A_{\theta}^{(i)}(r)\exp [i(\omega t-m\theta)]\
,
\end{eqnarray}
with $m$ as an integer for azimuthal variations and $i=s,\ g$ for
the stellar and magnetized gaseous discs respectively; parameter
$\omega$ is the angular frequency of coplanar perturbations and is
connected to the angular wave pattern speed $\Omega_p$ by
$\omega=m \Omega_p\ $.
The perturbation surface mass density is set to the following form
\begin{eqnarray}
\Sigma^{(i)}_1=S^{(i)}_1r^{-2\beta_1-1}\exp[i(\omega t-m\theta +\nu
\ln r)]
=S^{(i)}_1r^{-2\beta^d_1-1}\exp[i(\omega t-m\theta)]\
\end{eqnarray}
with
$\beta^d_1\equiv\beta_1-i\nu/2\ .$
The perturbation coefficient $S^{(i)}_1$ is a small amplitude
coefficient for each disc. The complex phase factors $\exp
(im\theta)$ and $\exp(i\nu\ln r)$ represent azimuthal and radial
variations, respectively. In general, index $\beta_1$ can be
different from $\beta$ of the equilibrium disc configuration. The
dispersion relation for coplanar perturbations in a
gravitationally coupled disc configuration is then given by
\begin{eqnarray}\label{eq:dr}
\big[(a^{(s)})^2-G\Sigma^{(s)}_0 r
Y_m(\beta^d_1)-\overline{\omega}^{(s)}r^2/Q^{(s)}\big]
\big[\Theta (a^{(g)})^2-\epsilon G\Sigma^{(g)}_0 r
Y_m(\beta^d_1)-\overline{\omega}^{(g)}r^2/Q^{(g)}\big]
=\big[G\Sigma^{(s)}_0 rY_m(\beta^d_1)\big]\big[G\Sigma^{(g)}_0 r
Y_m(\beta^d_1)\big]\ ,
\end{eqnarray}
with the notation abbreviation
$\overline{\omega}^{(i)}\equiv\omega-m\Omega^{(i)}_0$; and $Q^{(s)}$
and $Q^{(g)}$ are two abbreviations defined in Appendix
\ref{sec:dr}. Derivation details of this dispersion relation are
presented in Appendix \ref{sec:dr}. Dispersion relation
(\ref{eq:dr}) is of great importance and is very powerful as it
contains all useful information about disc dynamics in the presence
of coplanar perturbations. The regime of small $\omega$ corresponds
to the quasi-stationary situations for the QSSS hypothesis (e.g.
Bertin \& Lin 1996 and extensive references therein).
To be specific and for simplicity, we set $\omega=0$ in our frame of
reference to study the case of non-axisymmetric stationary
perturbations with $m\neq 0$ which can be either aligned or
unaligned. Dispersion relation (\ref{eq:dr}) can be reduced to a
stationary dispersion relation as
\begin{eqnarray}
\left\lbrace 1-\frac{GS^{(s)}_0Y_m(\beta^d_1)}{\big[a^{(s)}\big]^2
r^{2\beta}}-\big[D^{(s)}\big]^2\left[\frac{m^2-2(1-\beta
)}{m^2-4(\beta^d_1)^2+2\beta^d_1+2\beta}\right]\right\rbrace
\qquad\qquad\qquad\qquad\qquad\qquad\qquad\qquad\qquad\qquad\qquad\nonumber \\
\qquad\qquad\times\left\lbrace 1-\frac{\epsilon
GS^{(g)}_0Y_m(\beta^d_1)}{\Theta \big[a^{(g)}\big]^2
r^{2\beta}}-\big[D^{(g)}\big]^2\left[\frac{m^2-2(1-\beta)}{m^2-4
(\beta^d_1)^2+2\beta^d_1+2\beta }\right]\right\rbrace
=\left\lbrace\frac{\rmn{G}S^{(s)}_0Y_m(\beta^d_1)}{\big[a^{(s)}\big]^2
r^{2\beta}}\right\rbrace\left\lbrace\frac{GS^{(g)}_0
Y_m(\beta^d_1)}{\Theta\big[a^{(g)}\big]^2r^{2\beta}}\right\rbrace\
,\label{disp}
\end{eqnarray}
where $D^{(s)}$ and $D^{(g)}$ are respectively the two disc
rotational Mach numbers as defined in Appendix \ref{sec:dr}.

By using equations (\ref{eq:sss}), (\ref{eq:ssg}),
(\ref{eq:D^{(s)}}) and (\ref{eq:D^{(g)}}), stationary dispersion
relation (\ref{disp}) can be cast into the explicit form of
\begin{eqnarray}
\left\lbrace 1-\frac{(1+2\beta)G(S^{(g)}_0)^{{1}/{(1+2\beta)}}
Y_m(\beta^d_1)}{(1+4\beta)k^{(s)}\delta_0^{{1}/{(1+2\beta)}}}-\left[
\frac{G(S^{(g)}_0)^{{1}/{(1+2\beta)}}2\beta
Y_0(\beta)(1+\delta_0)(1+f)}{(1+4\beta)k^{(s)}
\delta_0^{{1}/{(1+2\beta)}}}-1\right]
\frac{(2\beta+1)\big[m^2-2(1-\beta )\big]}{m^2-4(\beta^d_1
)^2+2\beta^d_1+2\beta}
\right\rbrace\nonumber \\
\times\bigg\lbrace 1-\frac{(1+2\beta)\epsilon G
(S^{(g)}_0)^{{1}/{(1+2\beta)}}Y_m(\beta^d_1)}{(1+4\beta)\Theta
k^{(g)}} -\left[ \frac{2\beta G(S^{(g)}_0)^{{1}/{(1+2\beta)}}
Y_0(\beta) (f(1+\delta_0^{-1})+\delta_0^{-1}+\epsilon)}{(1+4\beta)
\Theta k^{(g)}}-1\right]\nonumber\\
\times\frac{(2\beta+1)\big[m^2-2(1-\beta )\big]}
{m^2-4(\beta^d_1)^2+2\beta^d_1+2\beta }\bigg\rbrace=
\frac{(1+2\beta)G(S^{(g)}_0)^{{1}/{(1+2\beta)}}
Y_m(\beta^d_1)}{(1+4\beta)k^{(s)}\delta_0^{{1}/{(1+2\beta)}}}
\frac{(1+2\beta)G(S^{(g)}_0)^{{1}/{(1+2\beta)}}
Y_m(\beta^d_1)}{(1+4\beta)\Theta k^{(g)}}\ .\label{eq:dimless}
\end{eqnarray}
Dimensionless parameters $\beta$, $m$ and $\beta_1^d$ can be
chosen within the allowed ranges. Dimensionless parameters
$\epsilon$ and $\Theta$ depend on the isopedic magnetic field
characterized by the dimensionless isopedic parameter $\lambda$.
For parameters $S_0^{(g)}$, $k^{(s)}$, $k^{(g)}$ and $\delta_0$,
typical values of late-type spiral galaxies are adopted and
described in details in Section \ref{sec:parameters} below.

\subsection{Relation between Dark Matter Halo and Isopedic Magnetic Field}

The goal of our investigation is to explore the functional
relation $f(\lambda)$ between the amount of dark matter and the
isopedic magnetic field with other parameters specified. For this
purpose, we further introduce several simplifying abbreviations
for the derivation of $f(\lambda)$ relation, viz.
\begin{eqnarray}
M\equiv\frac{m^2-2(1-\beta )}{m^2-4(\beta^d_1)^2+2
\beta^d_1+2\beta}\ ,\label{eq:M}
\qquad\qquad
B^{(s)}\equiv\frac{GS^{(s)}_0}{\big[a^{(s)}\big]^2r^{2\beta}}\
,\label{Bs}
\qquad\qquad
B^{(g)}\equiv\frac{GS^{(g)}_0}{\Theta\big[a^{(g)}\big]^2
r^{2\beta}}\ ,\label{Bg}
\end{eqnarray}
which are all dimensionless. The simplified form of stationary
dispersion relation (\ref{disp}) then appears as
\begin{eqnarray}\label{sdisp}
\left\lbrace
1-B^{(s)}Y_m(\beta^d_1)-\big[D^{(s)}\big]^2M\right\rbrace
\left\lbrace 1-\epsilon
B^{(g)}Y_m(\beta^d_1)-\big[D^{(g)}\big]^2M\right\rbrace
=\left[B^{(s)}
Y_m(\beta^d_1)\right]\left[B^{(g)}Y_m(\beta^d_1)\right]\ ,
\end{eqnarray}
where the two disc rotational Mach numbers squared are defined by
\begin{eqnarray}
\left[D^{(s)}\right]^2=2\beta B^{(s)}Y_0(\beta)
(1+\delta_0)(1+f)-2\beta-1\ ,\qquad\quad\ \\
\left[D^{(g)}\right]^2=2\beta
B^{(g)}Y_0(\beta)\big[f(1+\delta_0^{-1})
+\delta_0^{-1}+\epsilon\big]-2\beta-1\ .
\end{eqnarray}
The two parameters $M$ and $Y_m(\beta^d_1)$ are complex in general
as parameter $\beta_1^d\equiv\beta_1-i\nu/2$ is complex in
general. But for two separate special cases $\beta_1^d=\beta_1$
(i.e. $\nu=0$) and $\beta_1=1/4$ with $\nu\neq 0$, we have shown
in Appendix (\ref{sec:f,lambda}) that both $M$ and
$Y_m(\beta^d_1)$ become real numbers. In the literature,
the situation of $\nu=0$ is called the aligned case because
perturbation patterns appear aligned along the radial direction;
that is, only azimuthal variations but no radial variations are
involved, and the situation of $\nu\neq 0$ is called unaligned
case because perturbation patterns appear spiral-like.

>From now on, we shall only focus on these two special real cases
$\nu=0$ with $\beta_1^d=\beta_1=\beta$ and $\nu\neq 0$ with
$\beta_1=1/4$ in the remainder of this manuscript. Unless
otherwise stated, we have either $\beta_1^d=\beta$ or
$\beta_1^d=1/4-i\nu/2$ in all the following equations. With our
notations, stationary dispersion relation (\ref{sdisp}) for
non-axisymmetric stationary perturbations can be cast into the
following form of
\begin{eqnarray}\label{eq:eq}
[1-B^{(s)}Y_m(\beta^d_1)+(2\beta+1)M
-2\beta B^{(s)}Y_0(\beta)(1+\delta_0)(1+f)M]
\qquad\qquad\qquad\qquad\qquad\qquad\qquad\qquad\qquad\\
\nonumber\qquad\qquad\times\big\lbrace-2\beta
B^{(g)}Y_0(\beta)[f(1+\delta_0^{-1})+\delta_0^{-1}+\epsilon]M+1
-\epsilon B^{(g)}Y_m(\beta^d_1)+(2\beta+1)M
\big\rbrace=B^{(s)}B^{(g)}\big[Y_m(\beta^d_1)\big]^2\ .
\end{eqnarray}
In the absence of the gravitational coupling on the right-hand
side (RHS) of equation (\ref{eq:eq}), the first and second factors
on the left-hand side (LHS) represent stationary dispersion
relations in stellar and magnetized gaseous discs separately. That
is, for a single stellar disc, we can determine a $f$ value for
stationary perturbations. Or, for a single magnetized gaseous
disc, we can also determine a $f$ value for stationary MHD
perturbations. When the two discs are coupled by gravity, we need
to determine the $f$ parameter in a joint manner by dispersion
relation (\ref{eq:eq}) for stationary perturbation patterns in
both discs with different angular rotation speeds. Again, two
notational abbreviations are introduced below for the convenience
of further derivations, viz.
\begin{eqnarray}
C_1\equiv 1-2\beta B^{(s)}Y_0(\beta)(1+\delta_0)M
-B^{(s)}Y_m(\beta^d_1)+(2\beta+1)M\ ,\quad\\
C_2\equiv 1-2\beta B^{(g)} Y_0(\beta)(\epsilon+\delta_0^{-1})M
-\epsilon B^{(g)}Y_m(\beta^d_1)+(2\beta+1)M\ .
\end{eqnarray}
Then stationary dispersion relation (\ref{eq:eq}) for coplanar
perturbations can be rearranged into the form of
\begin{eqnarray}
\big[C_1-2\beta B^{(s)}Y_0(\beta)(1+\delta_0)M f\big]
\big[C_2-2\beta B^{(g)}
Y_0(\beta)f(1+\delta_0^{-1})M\big]-B^{(s)}B^{(g)}
\big[Y_m(\beta_1^d)\big]^2=0\ .
\end{eqnarray}
This leads to an explicit quadratic equation in terms of the
gravitational potential ratio $f$ given below
\begin{eqnarray}
f^2-f\frac{C_1B^{(g)}(1+\delta_0^{-1})+C_2B^{(s)}(1+\delta_0)}
{2\beta B^{(s)}B^{(g)}Y_0(\beta)M(1+\delta_0)(1+\delta_0^{-1})}
+\frac{C_1C_2-B^{(s)}B^{(g)}\big[Y_m(\beta_1^d)\big]^2}{4\beta^2B^{(s)}B^{(g)}
[Y_0(\beta)]^2M^2(1+\delta_0)(1+\delta_0^{-1})}=0\ .
\end{eqnarray}
By further introducing two notational abbreviations
\begin{eqnarray}
V^{(g)}\equiv B^{(g)}(1+\delta_0^{-1})\ ,
\hspace{2.5cm} V^{(s)}\equiv B^{(s)}(1+\delta_0)\ ,
\end{eqnarray}
the two roots $f_{1}$ and $f_{2}$ for the gravitational potential
ratio $f$ are simply
\begin{eqnarray}\label{roots}
f_{1,2}\equiv f_{\pm}=\frac{C_1V^{(g)}+C_2V^{(s)}}{4\beta
V^{(s)}V^{(g)}Y_0(\beta)M}
\pm\frac{\big\{\big[C_1 V^{(g)}-C_2 V^{(s)}\big]^2+4
\big[Y_m(\beta^d_1)\big]^2
V^{(s)}V^{(g)}B^{(s)}B^{(g)}\big\}^{1/2}}{4\beta
V^{(s)}V^{(g)}Y_0(\beta)M}\ ,
\end{eqnarray}
where the two roots $f_1\equiv f_{+}$ and $f_2\equiv f_{-}$
correspond to the plus and minus signs, respectively. All
variables are real numbers in our model study here for either
$\beta_1^d=\beta$ or $\beta_1^d=1/4-i\nu/2$, and there always
exist two roots for $f$ which can be either both real or a pair of
complex conjugates. By our definition of $f=\overline{\Phi}/\Phi$,
we should identify the real positive solution(s) for $f$ ratio.
This leads to the requirement that the determinant under the
square root in expression (\ref{roots}) must be positive which
turns out to be automatically satisfied.
Another requirement is that in equations (\ref{eq:omegastellar})
and (\ref{eq:omegagaseous}) of Appendix A, both $v_{r0}^2$ and
$v_{\theta 0}^2$ should be non-negative which leads to the
following conditions,
\begin{eqnarray}
f\geq \frac{(1+4\beta) k^{(s)}
\big[S^{(s)}_0\big]^{{2\beta}/{(1+2\beta)}}}{2\beta G
Y_0(\beta)S_0}-1\hspace{1 cm}\rmn{and}\hspace{1 cm}
f\geq\frac{\Theta (1+4\beta)k^{(g)}
\big[S^{(g)}_0\big]^{{2\beta}/{(1+2\beta)}}}{2\beta G
Y_0(\beta)S_0}+\frac{(1-\epsilon)S^{(g)}_0}{S_0}-1\ .
\label{eq:fanforderung}
\end{eqnarray}
Together with inequality (\ref{eq:f(delta)}), there are thus total
four necessary conditions on gravitational potential ratio $f$,
viz.
\begin{itemize}
\item a)\hskip 0.5cm  $f \in \Re_+$\ , \item b)\hskip 0.5cm
$\Sigma^{(g)}_0(f)>0$\ , \item c)\hskip 0.5cm
$\big[v^{(s)}_{\theta 0}\big]^2(f) >0$\ , \item d)\hskip 0.5cm
$\big[v^{(g)}_{\theta 0}\big]^2(f)>0$\ ,
\end{itemize}
in our model analysis for global stationary perturbation patterns.

\section{Parameters for magnetized spiral galaxies}\label{sec:parameters}

In dispersion relation (\ref{eq:dimless}), dimensionless
parameters $m$, $\beta$, $\beta_1^d$, $\epsilon$ and $\Theta$
should be specified to characterize stationary perturbations in
composite disc system. Physically, integer $m$ indicates the
number of spiral arms for a perturbation pattern, $\beta$ is the
power-law exponent to characterize radial variations of the
different unperturbed background variables, $\beta_1^d$ is the
exponent to characterize radial variations of perturbations.
Parameters $\epsilon$ and $\Theta$ are the reduction and
enhancement factors associated with an isopedic magnetic field,
respectively. For relevant coefficient $S_0^{(g)}$, we adopt
parameter ranges listed in Table \ref{tab:late-typeparameters} in
reference to estimates from observations (e.g. Roberts 1962).
\begin{table*}
\caption{Parameters for the study of late-type disc galaxies with
isopedic magnetic field. Numerical ranges for $S^{(g)}_0$ and
$S^{(s)}_0$ are calculated for $\beta=[0,\ 0.5]$; the ranges for
$\lambda$ and $\epsilon$ are determined with the given ranges of
$\Sigma_0^g$ and $B_z$ tabulated below. }
\begin{tabular}{@{}ll@{}}
\hline
Variables & Value \\
\hline
$\Sigma_0^g$ & $2\times 10^{-4}-2\times 10^{-3}\ \rmn{g\ cm}^{-2}$ \\
\hline
 $\Sigma^{(g)}_0 (5\,\rmn{kpc})$ & $10^{-3}\ \rmn{g\ cm}^{-2}$ \\
\hline
$S^{(g)}_0=\Sigma^{(g)}_0 r^{2\beta+1}$ & $(10^{19} ,\ 10^{41})\
\rmn{g\ cm}^{2\beta-1}$ \\
\hline
$\delta_0=\Sigma^{(g)}_0/\Sigma^{(s)}_0$ & 0.05 \\
\hline
$a^{(g)}(5\,\rmn{kpc})$  & $1 \times 10^{5}\, \rmn{cm\ s}^{-1}$  \\
\hline
$a^{(s)}(5\,\rmn{kpc})$\footnote{corresponding to the stellar velocity
dispersion} & $3\times 10^{6}\,\rmn{cm\ s}^{-1}$ \\
\hline
$B_z$ & $(1,\ 10)\ \rmn{\mu G}$\\
\hline
$B_z$ in central regions & $(20,\ 40)\ \rmn{\mu G}$\\
\hline
$\lambda$ & $(0.03,\ 3.25)$ \\
\hline
$\epsilon$& $(-1110,\ 1)$\\
\hline
\end{tabular}
\label{tab:late-typeparameters}
\end{table*}
Two parameters $k^{(s)}$ and $k^{(g)}$ are functions of $\beta$
parameter and closely relate  to the sound speeds (as given in
Table \ref{tab:late-typeparameters}) according to equations
(\ref{eq:sss}) and (\ref{eq:ssg}) with cgs unit of
$\rmn{cm^{2n}/(s^2\ g^{n-1})}$. Typical values for $k^{(s)}$ and
$k^{(g)}$ are of the order $\sim 10^{10}\ \rmn{cm^{2n}/(s^2\
g^{n-1})}$. The range for $\Theta$ is a function of $\beta$ and
$\delta_0$ and cannot be given here simply but can be computed in
a straightforward manner. For the Gamma functions $\Gamma (z)$
involved in our computations, we use the code developed by Zhang
\& Jin (1996).\footnote{For Gamma functions $\Gamma (z)$ in C/C++
languages of real and complex arguments, the reader is referred to
website http://www.crbond.com/math.htm\ .} Our study explores
behaviours of potential ratio $f$ dependence on isopedic parameter
$\lambda$ for globally aligned and unaligned logarithmic
stationary perturbation patterns. Physically, this study reveals
possible relationships of an isopedic magnetic field and the
gravitational potential of an axisymmetric dark matter halo for
stationary global perturbations. Since
$Y_m(\beta^d_1)=Y_{-m}(\beta^d_1)$ (see footnote\footnote{A
mathematical proof of this relation can be found in Appendix A of
Wu \& Lou (2006).}) and $M(m^2)=M((-m)^2)$, we focus on the case
of $m>0$ without any loss of generality. We present only the cases
for $m=1$ and $m=2$ since a previous analysis of Lou \& Wu (2005)
has shown that all curves for $m\geq 2$ bear the similar form,
whereas the case of $m=1$ carries a unique form.

\section{Global Stationary Aligned Perturbation Patterns
with $\nu=0$ and $\beta_1=\beta$}\label{sec:nu=0}

\begin{figure}
\begin{center}
\begin{tabular}{cc}
\resizebox{70mm}{!}{\includegraphics{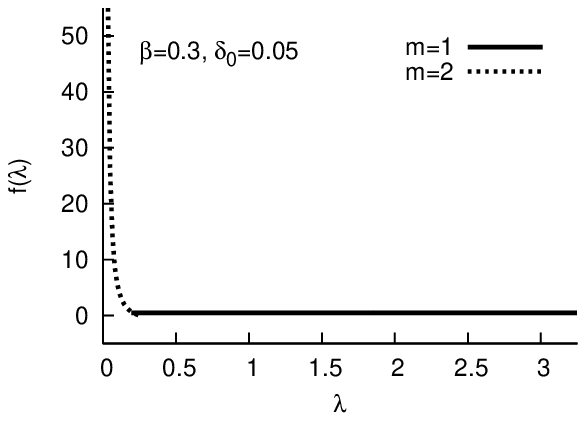}} &
\resizebox{70mm}{!}{\includegraphics{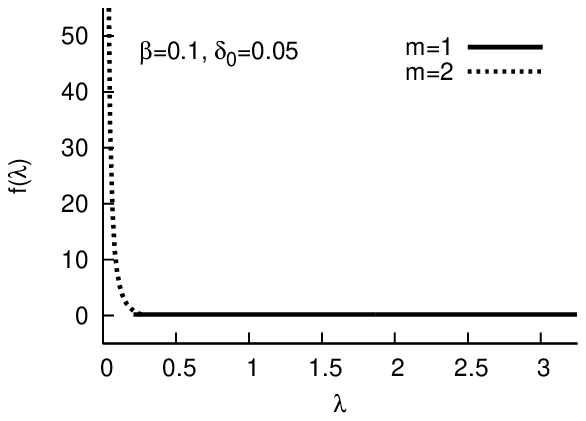}} \\
\end{tabular}
\caption{The dependence of the gravitational potential ratio $f$
on the dimensionless ratio $\lambda=2\pi
G^{1/2}\Sigma_0^{(g)}/B_z$ and $m$ in the stationary aligned case
for late-type disc galaxies with $\delta_0=0.05$. Different
$\beta$ values are studied for the aligned case of $\nu=0$ and
$\beta_1=\beta$.
For this and all the following figures,
values of all other parameters, which are not explicitly
mentioned, are listed in Table \ref{tab:late-typeparameters} of
Section \ref{sec:parameters}. For $\beta=0.1$ and $m=1$ (right
panel), $f\rightarrow\sim 0.2$, while for $\beta=0.3$ and $m=1$
(left panel), $f\rightarrow\sim 0.5$.
}\label{fig:betachange}
\end{center}
\end{figure}

\begin{figure}
\begin{center}
\begin{tabular}{cc}
\resizebox{70mm}{!}{\includegraphics{b=0.3,d=0.05.eps}} &
\resizebox{70mm}{!}{\includegraphics{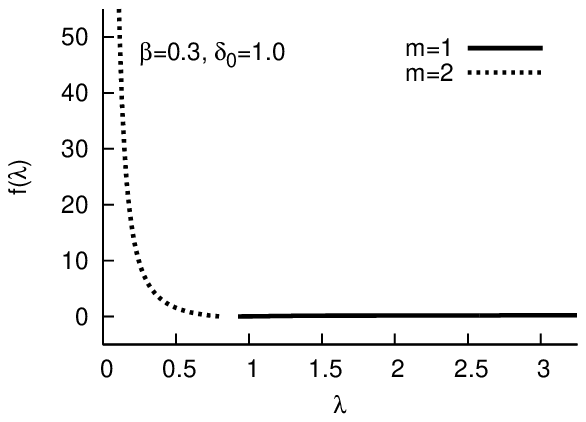}} \\
\resizebox{70mm}{!}{\includegraphics{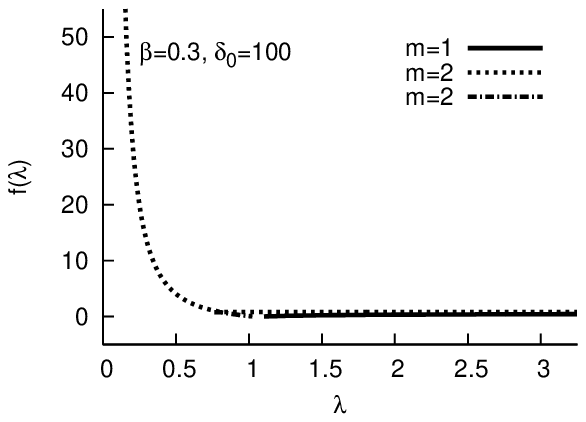}} &
\resizebox{70mm}{!}{\includegraphics{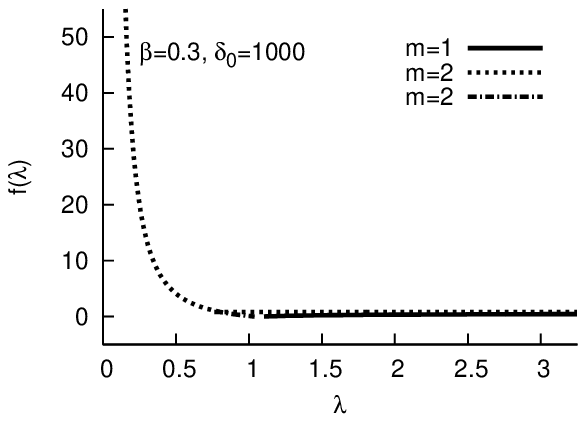}}\\
\end{tabular}
\caption{The aligned case with $\nu=0$ and $\beta_1=\beta$ is
studied. The dependence of the dark matter amount $f$ on the
dimensionless ratio $\lambda$ and $m$ in the stationary case for a
spiral galaxy with $\beta=0.3$ and different $\delta_0$
corresponding to different epochs of a spiral galaxy evolution. }
\label{fig:deltachange}
\end{center}
\end{figure}

For global stationary aligned cases with $\nu=0$, we have further
taken a special case of $\beta^d_1=\beta_1=\beta$ for
perturbations carrying the same scale-free index as that of the
background equilibrium disc configuration. Relations $f(\lambda)$
for different combinations of $\beta$ and $\delta_0$ are shown in
Figures \ref{fig:betachange} and \ref{fig:deltachange} as
examples.
We note first that no sensible $f_2$ root can be found in Figure
\ref{fig:betachange} satisfying requirements c) and d) listed
after inequalities (\ref{eq:fanforderung}).
In the following, we only show stationary perturbation solutions
with $f$ satisfying all four requirements a)$-$d) above.

We have explored
$\beta$ values within the range $\beta=[0.01,0.49]$. Figure
\ref{fig:betachange} shows examples for two different $\beta$
values. The numerical results for $\beta>0$ values that we have
studied do not show remarkable differences among all $m$ values.
While $m\geq 2$ give monotonically decreasing functions
$f(\lambda)$ for all $\beta>0$, $m=1$ case leads to an almost
constant $f(\lambda)$ whose value decreases with decreasing
$\beta$. This latter trend of decrease cannot be easily discerned
in the figures shown, but can be readily found by checking the
specific numbers. For $m\geq 2$, stationary solutions for aligned
perturbations can be found in all figures for small $\lambda$
corresponding to stronger magnetic fields. For $m=1$, stationary
solutions for aligned perturbations exist in a wide range of
$\lambda$ values. In general, $f=10$ is a typical value for spiral
galaxies containing dark matter halos. In Figure
\ref{fig:betachange} of a late-type galaxy, $f=10$ leads to
$\lambda\cong 0.1$. For a gaseous surface mass density of
$\Sigma^{(g)}_0=2 \times 10^{-4}\ \rmn{g\ cm}^{-2}$, one gets
$B_z=4\ \rmn{\mu G}$ which is a realistic number and for
$\Sigma^{(g)}_0=2\times 10^{-3}\ \rmn{g\ cm}^{-2}$, the magnetic
field strength is $B_z=40\,\rmn{\mu G}$ which may be reached in
the central regions of spiral galaxies. It is then realistic to
have stationary aligned $m=1$ perturbations corresponding to
lopsided patterns.

An increase of $\delta_0$ in Figure \ref{fig:deltachange} should
be understood as going back to earlier evolution phases of a
spiral galaxy. This is because at early times, the gaseous disc
contains more mass and during the course of evolution, the stellar
disc becomes more and more massive as a result of star formation.
When $\delta_0$ is increased, the stationary range for $m\geq 2$
is also increased to larger $\lambda$ values whereas that of $m=1$
is decreased. For larger values of $\delta_0$, $m\geq 2$ gives
stationary solutions in the entire range of $\lambda$. In
addition, a second solution of $m=2$ can be also found in the
range around $\lambda\in (0.7\sim 1)$ for the two larger
$\delta_0$ values. In the limit of $\delta_0\to\infty$, the
stellar disc contains a negligible amount of mass as compared to
that of the magnetized gaseous disc. Effectively, this situation
can be understood as the presence of only one single isopedically
magnetized gaseous disc which was investigated earlier by Wu \&
Lou (2006). For $\delta_0>0.05$, we study again the case of
$f=10$. Here, $\lambda$ can be much larger than only 0.1 for
late-type spiral galaxies. For $\delta_0=1000$, one gets
$\lambda=0.4$. In this case, magnetic fields are $B_z=1\rmn{\mu
G}$ for $\Sigma^{(g)}_0=2\times 10^{-4}\rmn{g\ cm}^{-2}$ which is
also realistic and $B_z=10\rmn{\mu G}$ for $\Sigma^{(g)}_0=2\times
10^{-3}\rmn{g\ cm}^{-2}$.

The case of $m=1$ represents an exceptional case in our
exploration so far. Observational evidence for the existence of
such so-called lopsided galaxies can be found in Baldwin,
Lynden-Bell \& Sancisi (1980). For a late-type spiral galaxy with
$\delta_0\sim 0.05$ and a positive $\beta$ value, the stationary
solutions in Figure 1 can be roughly divided into two ranges of
$\lambda$. On the left side with small $\lambda$ or strong
magnetic field strengths, stationary solutions for $m\geq 2$ are
found. In almost the entire range of $0.2\leq\lambda\leq 3.25$,
the $m=1$ case has stationary solutions. In general, in the range
of weak magnetic fields only $m=1$ has stationary solutions. For
strong magnetic fields, only $m\geq 2$ can possess stationary
solutions. The limit of $\lambda\to\infty$ can be regarded as the
absence of magnetic fields which was analyzed by Shen \& Lou
(2004).

For larger $\delta_0$ corresponding to an earlier phase of an
evolving spiral galaxy as shown in Figure 2, the stationary range
of $\lambda$ for $m=1$ shrinks whereas that of $m\geq 2$ increases
to all possible physical values of $\lambda$; within a certain
range of $\lambda$, a second solution for $m\geq 2$ also comes
into existence.

Independent of parameter variations, $f(\lambda)$ remains always
almost constant for large $\lambda$ or weak magnetic field
strengths. This is sensible as weak magnetic fields are not
expected to affect the entire perturbation configuration
significantly (e.g. Wentzel 1963). For strong magnetic fields, we
need to have a larger amount of dark matter to maintain stationary
perturbation configurations. In the aligned case, this increase of
$f$ can reach values of several tens, e.g. in the limit of
$\lambda=0.03$, we have $f\sim 45,\ 55,\ 75$ for $\beta=0.49,\
0.3,\ 0.1$, respectively. The exploration of different $\delta_0$
values has revealed additional information: in disc evolution in
the early Universe, magnetic field should have played a more
important role in a spiral galaxy. This is because the range of
$\lambda$ with a significant change of $f$ appears larger; thus
weaker magnetic fields should have also affected $f$ values. Note
that the magnetic field cannot cause a growth of dark matter halo
in a spiral galaxy. Nevertheless, this analysis reveals the
relation between the dark matter ($f$ ratio) and the isopedic
magnetic field strength ($\lambda$ parameter). As expected for
weak magnetic fields, one may ignore $B_z$ in general. While for
strong magnetic fields, there is only one possible $f$ for each
$\lambda$ in order to maintain globally stationary perturbation
configurations. Physically, a larger $f$ ratio also leads to a
larger disc rotation speed $v_{\theta 0}^{(i)}$ according to
equations (\ref{eq:omegastellar}) and (\ref{eq:omegagaseous}). For
example, one estimates $v^{(s)}_{\theta 0}\cong 270\ \rmn{km\
s}^{-1}$ for $f=20$, $r=10\ \rmn{kpc}$, $\beta=0.1$, $a^{(s)}=30\
\rmn{km\ s}^{-1}$, $\Sigma^{(g)}_0=10^{-4}\ \rmn{g\ cm}^{-2}$ and
$\delta_0=0.05$ for a late-type spiral galaxy; while for $f=30$
with other parameters the same, one obtains $v^{(s)}_{\theta
0}\cong 330\ \rmn{km\ s}^{-1}$. Although very large disc
rotational velocities have not been observed, the part with large
$f$ can be regarded as theoretically plausible solutions for
spiral galaxies.

\section{Global Stationary Unaligned Logarithmic
Spiral Configurations with $\nu\neq 0$ and
$\beta_1=1/4$}\label{sec:nu!=0}

When MHD density wave perturbations propagate in both radial and
azimuthal directions, we have global unaligned logarithmic spiral
patterns with $\nu\neq 0$
and $m\neq 0$. In our consideration here, we take
$\beta^d_1=\beta_1-i\nu/2\neq\beta$ for $\nu\neq 0$. In this case,
$M$ and $Y_0(\beta_1^d)$ as defined by equations (\ref{eq:M}) and
(\ref{Y0def}) are complex in general, while dimensionless
parameters $B^{(s)}$ and $B^{(g)}$ as defined by equations
(\ref{Bs})
are always real. In the special case of $\beta_1=1/4$, we have shown
that both $M$ and $Y_0(\beta_1^d)$ are real (see Appendix A).
Physically, this case of $\beta_1=1/4$ corresponds to a constant
radial flux of angular momentum (e.g. Goldreich \& Tremaine 1979).
Again, the stationary condition for logarithmic spiral patterns
implies a certain relation between $f$ ratio (related to the dark
matter halo) and $\lambda$ parameter (related to the isopedic
magnetic field). For $\nu\neq 0$, a special attention is further
paid to $\nu-$dependence of $f(\lambda)$ (see footnote 3).
For different $\nu$ values
(related to the radial wave number) in a warm disc, the curves
$f(\lambda)$ are shown in Figure \ref{fig:f(Lambda,m)beta_1=1/4}.
\begin{figure}
\begin{center}
\begin{tabular}{cc}
\resizebox{70mm}{!}{\includegraphics{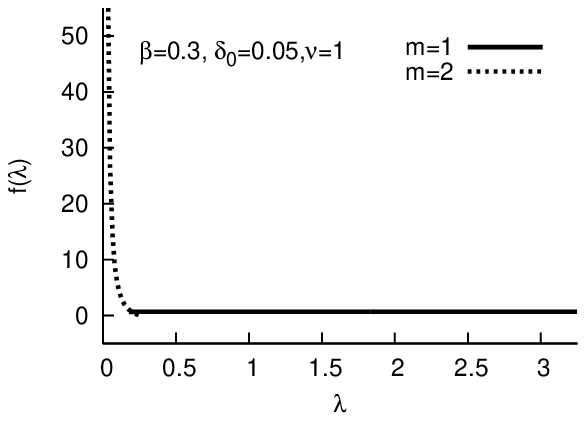}} &
\resizebox{70mm}{!}{\includegraphics{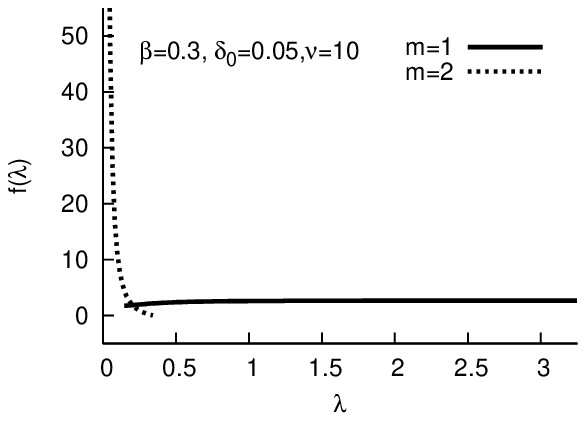}} \\
\resizebox{70mm}{!}{\includegraphics{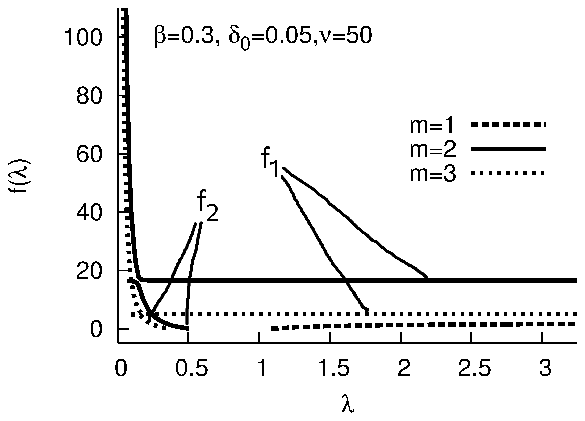}} &
\resizebox{70mm}{!}{\includegraphics{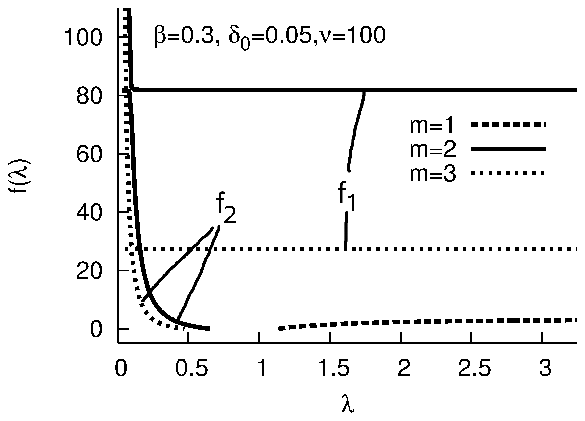}}\\
\end{tabular}
\caption{The function $f(\lambda,\ m)$ for $\beta=0.3$,
$\delta_0=0.05$ and $\beta_1=1/4$ in the unaligned logarithmic case with
$\beta_1=1/4$.
In the lower two panels, the two branches of $f_1$ and $f_2$ with
$f_1>f_2$ are shown for each $m\geq 2$ (see Appendix B). Note that
for small $\lambda$, $f_1$ and $f_2$ branches do not cross each
other, although they appear to be extremely close.
}\label{fig:f(Lambda,m)beta_1=1/4}
\end{center}
\end{figure}
For small $\nu$, at least one stationary solution exists for each
$m>0$. However for increasing $\nu$ towards the tight-winding
regime, stationary perturbation solutions of $m=1$ are shifted
towards weaker magnetic fields. For $\nu=50$, solutions for $m=1$
are found within the range $\lambda\approx (1.1,3.25)$ and for
$\nu=100$, they are within $\lambda\approx (1.2,3.25)$. Meanwhile,
stationary perturbation solutions for $m\geq 2$ grow into a larger
range of $\lambda$. If $\nu$ is increased further (i.e. towards
the extremely tight-winding regime), two stationary perturbation
solutions exist corresponding to each $m$. Thus according to
expression (\ref{roots}) for the two roots of $f$, $f_1$ and $f_2$
are the upper and lower curves in Figure
\ref{fig:f(Lambda,m)beta_1=1/4}, respectively. In the range of
$\lambda$ for stationary perturbation solutions, $f_1$ remains
more or less constant so that a change of magnetic field strength
appears independent of the amount of dark matter. Meanwhile, $f_2$
changes over a larger range of $\lambda$, indicating that the
magnetic field and $f_2$ relate each other more closely.
The magnetic field is directly connected to $\lambda$. If $f_1$
remains constant for all $\lambda$, then the reverse direction of
reasoning cannot be used, i.e., for a given $f_1=19$ and $\nu=50$,
the exact value of $\lambda$ and so $B_z$ cannot be determined
with certainty. In contrast to $f_1$, $f_2$ varies for $\nu=50$
over a larger range of $\lambda=(0.1,0.5)$. In this range of
$\lambda$, one can determine for each $f_2$ the associated
$\lambda$ and thus $B_z$ through $B_z=2\pi\sqrt{G}
\Sigma^g_0/\lambda$.
The tighter the spiral arms are wound, the more dark matter is
needed in order to maintain stationary logarithmic spiral
patterns.
This is a general trend of variation and is valid for both roots
of $f$. Nevertheless, there exists a certain $\nu_0$ value such
that the second solution $f_2$ appears with a much less amount of
dark matter for $\nu>\nu_0$. In other words, with increasing $\nu$
the dark matter amount also increases, except for the second
solution with $\nu>\nu_0$. Physically,
the two roots $f_1$ and $f_2$ are equally valid.
But compared with the observational value of $f\approx 10$ for
typical spiral galaxies, $f_1$ appears too large, e.g. for
$\nu=50$ and $f_1\approx 19$ within the range of $\lambda=(0.2,\
3.25)$.
In comparison, the lower $f_2$ appears more plausible because of
its smaller values instead of the very large values of $f_1$.
For $\nu=100$, one can again find a relation between the $f_2$
curve and the magnetic field (represented by $\lambda$) and it is
thus also possible to determine $B_z$ through the $f$ ratio. If
the curve $f_2(\lambda)$ is sufficiently steep, so that a
one-to-one relation is present between $f_2$ and $\lambda$, then
the magnetic field strength $B_z$ can be determined by integral
(\ref{eq:lambda1}) into
\begin{equation}
 B_z=\frac{2\pi\sqrt{G}\Sigma^g_0}{\lambda}\ .
\end{equation}
The only additional variable that one needs to know is the surface
mass density of the magnetized gas disc.
%
In this paper, we have not explored situations of $\nu\neq 0$ and
$\beta_1\neq 1/4$ for which both parameters $M$ and
$Y_m(\beta_1-i\nu/2)$ are complex. In this case, an additional
study is needed to explore the variation of $\beta_1$ and its
influence on the gravitational potential ratio $f$. Physically,
$\beta_1$ is the constant scale-free index of surface mass density
perturbations. A larger $\beta_1$ corresponds to a steeper radial
profile of stationary coplanar perturbations in magnitudes.

\section{Model Results for Two Limiting Cases}\label{sec:limits}

In the first limit of large $\lambda$ regime (i.e. $B_z\rightarrow
0$), we readily reproduce the model results of Shen \& Lou (2004)
where the gravitational coupling of two scale-free stellar and
gaseous discs in the absence of magnetic field is analyzed.
\begin{figure}
\begin{center}
\begin{tabular}{cc}
\resizebox{70mm}{!}{\includegraphics{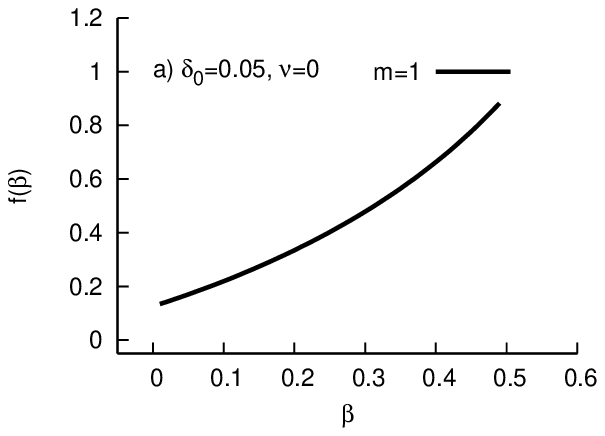}} &
\resizebox{70mm}{!}{\includegraphics{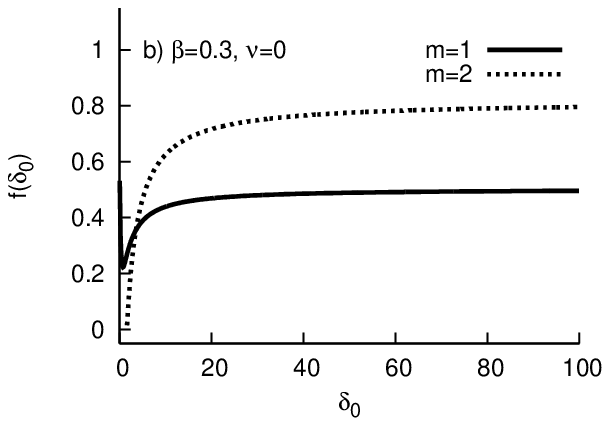}} \\
\resizebox{70mm}{!}{\includegraphics{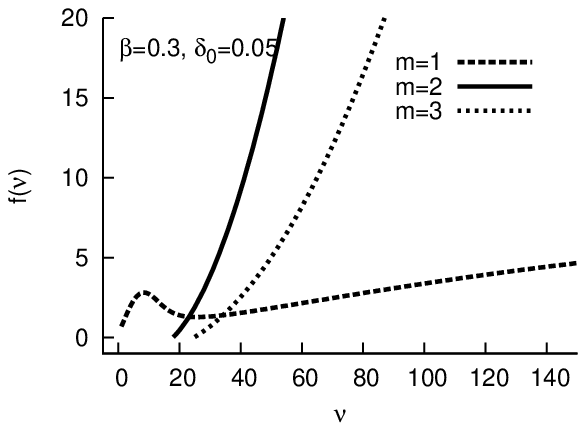}} \\
\end{tabular}
\caption{The limiting case of no magnetic field is approximately
realized by setting a large isopedic parameter $\lambda=24$.
} \label{fig:vergleichshen}
\end{center}
\end{figure}
For the absence of magnetic field, we choose a sufficiently large
isopedic parameter $\lambda=24$. The dependence of ratio $f$ on
various parameters is shown in Figure \ref{fig:vergleichshen}. For
the choice of $\delta_0=0.05$ and $\nu=0$, only the case of $m=1$
has stationary perturbation solutions. The variation of $\delta_0$
enables $m\geq 2$ to have stationary perturbation configurations
as already mentioned in Section \ref{sec:nu=0}. For $\nu=0$, each
$m$ corresponds to only one stationary perturbation solution. When
$\nu$ is sufficiently small, only $m=1$ perturbation
configurations have stationary solutions. As $\nu$ is gradually
increased, one can see clearly in Fig. \ref{fig:vergleichshen}
panel c) that stationary perturbation solutions for $m\geq 2$ also
appear.

\begin{figure}
\begin{center}
\begin{tabular}{cc}
\resizebox{70mm}{!}{\includegraphics{b=0.3,d=1000.eps}} &
\resizebox{70mm}{!}{\includegraphics{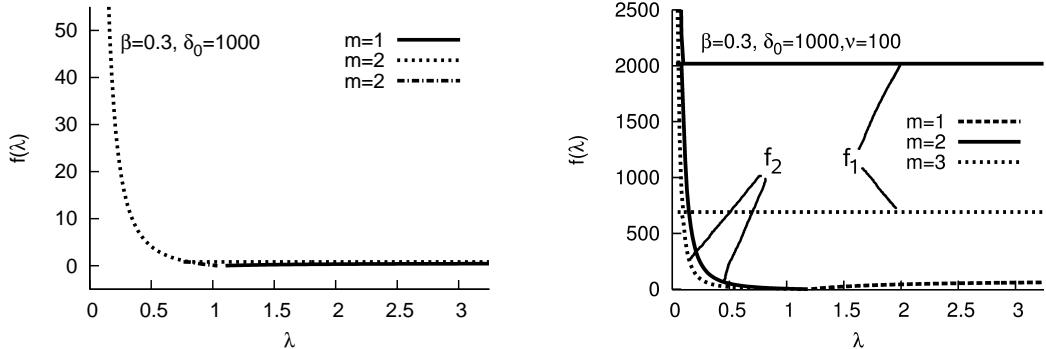}} \\
\end{tabular}
\caption{The limiting case of only one single isopedically
magnetized gaseous disc is approximately achieved by setting
$\delta_0=1000$.
Two branches $f_1$ and $f_2$ roots are shown with $f_1>f_2$ for
each $m\geq 2$ (see Appendix B). While they appear very close,
$f_1$ curve does not intersect $f_2$ curve at small $\lambda$.
} \label{fig:vergleichwu}
\end{center}
\end{figure}

In the second limit of large $\delta_0$ (i.e.
$\Sigma_0^{(s)}\rightarrow 0$) for young galaxies in the early
Universe, we recover the results of Lou \& Wu (2005) where global
MHD perturbation configurations in a single isopedically
magnetized gaseous disc are studied. Figure \ref{fig:vergleichwu}
shows different variations of $f$ ratio for the aligned and
unaligned logarithmic cases in that regime.
In principle, all $m$ values can have corresponding stationary
solutions for global perturbation configurations. This depends on
the choice of $\lambda$ and the combination of different
parameters such as $k^{(s)}$, $k^{(g)}$ or $S^{(g)}$ and so forth.
For $m=1$, the starting point of stationary solutions for the
aligned case is $\lambda\approx 1.1$ and for the unaligned case it
is slightly shifted to $\lambda\approx 1.2$. For $m=1$ the range
of ratio $f=(0.01,\ 0.5)$ is very small for $\nu=0$ as compared to
range $f=(1,\ 60)$ for $\nu=100$.
For $m\geq 2$, a variation of
$\beta$ parameter 
in the range of $\beta=(0.1,\ 0.49)$
for the background composite disc configuration does not affect
features of $f(\lambda)$ curves very much. In the model of Wu \&
Lou (2006), potential ratio $f$ is studied for different
properties of the magnetized gaseous disc alone (such as the sound
speed) at a certain specified value of $\lambda$ parameter. The
special case of one single magnetized gaseous disc in our model is
complementary to their analysis.
Again, the basic fact that typically $f\lsim 10$ must be kept in
mind for applications to spiral galaxies. Due to their smaller
values for $\nu=100$, the minus-sign solution $f_2$ is regarded as
more plausible.

\section{Conclusions and Discussion} \label{sec:summary}

We have investigated global stationary solutions for aligned and
unaligned logarithmic perturbation configurations with a constant
radial flux of angular momentum and paid special attention to the
roles of dark matter halo represented by the $f$ parameter for the
gravitational potential ratio and the isopedic parameter
$\lambda$. For this purpose, the stationary dispersion relation
for global perturbations is derived and a quadratic equation of
$f$ is obtained.
In our model formulation, the stationary
assumption is applied as a very special limiting case of the QSSS
hypothesis (e.g. Bertin \& Lin 1996 and references therein) which
itself is already a very strong requirement to the disc dynamics.
For a non-vanishing small pattern speed of a few $\mathrm{km\
s^{-1} kpc^{-1}}$ (adopted in the QSSS theory), a similar relation
between $f$ and $B_z$ may be also derived.\footnote{Wu \& Lou
(2009 in preparation) obtained a dispersion relation for non-zero
but small pattern speed and in the absence of the stellar disc
component (see also our expression (\ref{eq:dr}) and Wu \& Lou
2006 for very small angular perturbation frequency $\omega$).
The
range of potential ratio $f$ should be determined in order to
evaluate the importance of the isopedic magnetic field.
Intuitively, the results for quasi-stationary and stationary
configurations are expected to be qualitatively similar.
Quantitative deviations are expected to be proportional to certain
powers of pattern speed (e.g. Wu \& Lou 2006). } Our sample
calculations show that there are three possibilities for potential
ratio $f$, corresponding to no solution, one solution and two
solutions, respectively. In general, potential ratio $f$ is a
fairly complicated function of several independent model
parameters involved. In order to explore the dependence of
potential ratio $f$ upon $\beta$, $\delta_0$, $\lambda$, $m$ and
$\nu$, we have chosen typical values for these relevant parameters
to characterize late-type disc galaxies. Therefore, these results
should be applicable to typical spiral galaxies for comparison.
The main results are now summarized below.

\subsection{Global Aligned Perturbation Configurations
 with $\nu=0$ and $\beta_1=\beta$}

Global {\it aligned} perturbation configurations with $\nu=0$ are
studied for typical late-type galaxies with different $\beta$
values, corresponding to various power-law radial fall-offs of the
disc rotation curve. Examples of our numerical exploration show
that in the estimated range of $\lambda$ for disc galaxies, a
variation of $\beta$ parameter has no significant impact on global
stationary perturbation configurations with $m\geq 1$. As a
general trend, a stronger magnetic field allows existence of
stationary perturbation configurations for $m\geq 2$. For weaker
magnetic fields, only stationary perturbation configurations with
$m=1$ (i.e. lopsided cases) may exist. We have also explored
stationary perturbation configurations for a disc galaxy at
different epochs of evolution by varying the disc surface mass
density ratio $\delta_0=\Sigma^{(g)}_0/\Sigma^{(s)}_0$. By our
sample calculations, the variation of $\delta_0$ strongly affects
the behaviour of $f(\lambda)$. This is a consequence of the fact
that the magnetic field directly affects the gaseous disc and in
an earlier stage of galaxy evolution, the gas fraction is higher
in a composite disc system. With increasing $\delta_0$ ratio
towards earlier epochs (
i.e. at higher cosmological redshifts in the early Universe),
the range of allowed stationary
perturbation configurations for $m\geq 2$ also increases while
that for $m=1$ shrinks. In our simple scenario, $\delta_0$
parameter marks the evolution of a disc galaxy in the expanding
Universe. At the beginning, a proto-spiral galaxy is presumably
composed of a nebulous gas disc and a dark matter halo (e.g. Lou
\& Wu 2005; Wu \& Lou 2006). During the evolution of a spiral
galaxy, more and more stars are born and die and thus the disc
mass ratio $\delta_0$ decreases slowly with time. With this
scenario in mind, we interpret Figure \ref{fig:deltachange} as
follows. At an earlier stage of disc galaxy evolution, stationary
solutions of global perturbation configurations with $m\geq 2$
exist over a wider range of $\lambda$. The influence of magnetic
fields there is also stronger as the curve $f(\lambda)$ there
varies significantly over a larger range of $\lambda$, indicating
a wider range for magnetic field strengths.

This brings out an extremely interesting evolutionary
perspective for speculations. Observationally, the global star
formation rate (SFR) in spiral galaxies is an important indicative
parameter to characterize the galactic evolution. Conceptually, if
one attempts to relate this SFR with disc instabilities, the
application of Toomre's criterion (Safonov 1960; Toomre 1964) to a
stellar disc alone would be insufficient because stars are
directly born in the gaseous disc. Following this line of
reasoning, one needs at least to explore instabilities in a
composite system of stellar and gas discs (e.g. Lou \& Fan 1998,
2000 and references therein) in the presence of a massive dark
matter halo. The importance of magnetic fields is generally
recognized in the dynamics of star-forming cloud cores.
Nevertheless, how to specifically relate physical processes of
star formation in clouds on much smaller scales and of large-scale
disc instabilities remains a challenging problem due to the
tremendous differences in scales. In spite of this challenge, we
have developed intuitive feelings that regions of high gas density
and strong magnetic fields on large scales are expected to be
vulnerable or favorable to active star formation processes. Now in
a highly simplified dynamic manner, our model analysis brings
together several important aspects for this physical
consideration, viz. dark matter halo, magnetic field, and higher
gas fraction (i.e. larger $\delta_0$) in a composite disc system.
The initial conditions for forming a proto-galactic disc nebula
such as the dark matter halo, gas disc and magnetic field are
expected to be statistical with fluctuations. The instability
properties of such a magnetized disc system in the presence of a
dark matter halo may then grossly determine the global SFR and
thus the galactic evolution. In other words, different initial
conditions can lead to different kinds of evolutionary tracks. In
particular, it is conceivable that the SFR may have highs and lows
along galactic evolutionary paths in the expanding Universe. This
information would be valuable for understanding the overall
cosmological evolution in terms of global SFRs of spiral galaxies.

To compare a late-type spiral galaxy with an early-type spiral
galaxy, we emphasize that stationary perturbation solutions for
$m\geq 2$ are very different in these galaxies. In a late-type
spiral galaxy of $\delta_0\sim 0.05$, stationary perturbation
configurations of $m\geq 2$ can only exist for strong magnetic
fields while in an earlier stage of evolution, such solutions can
exist for almost all $\lambda$ values. These results suggest that
multiple-armed disc galaxies may be more numerous in the early
Universe. It may be possible for disc galaxies to change patterns
during the course of their evolution. Conceivably evolving on
cosmological timescales, a disc galaxy pattern may alternately
become stationary or quasi-stationary
as evidenced by the fact of changing $\delta_0$ and $\lambda$,
e.g. $f$ and $m\geq 2$ are shifted towards smaller $\lambda$ for
stationary configurations.
This offers a novel perspective to
relate global SFRs of disc galaxies and the galactic pattern speed
evolution. For example in numerical simulations, one may start
from initially stationary perturbation configurations and then
explore time-dependent quasi-stationary behaviours as well as
nonlinear effects of a composite disc system by adjusting one or
several relevant parameters systematically.

\subsection{Global Stationary Unaligned Logarithmic Spiral
Configurations with $\beta_1=1/4$ and $\nu\neq 0$}

Parameter $\nu$ characterizing radial variations of coplanar
perturbations affects $f$ ratio in many ways. For example, in
spiral galaxies with more tightly wound logarithmic spiral arms
(i.e. larger $\nu$), more dark matter is needed in the massive
halo to sustain stationary perturbation configurations.
Furthermore, with increasing $\nu$ towards the WKBJ regime, each
$f$ may correspond to multiple stationary configurations. For very
small $\nu$ in the opposite limit, only perturbation
configurations with $m=1$ have stationary solutions (i.e. lopsided
configurations).
Stationary perturbations with $m=1$ also have solutions for larger
$\nu$, but only in a certain range of $\lambda\approx (1.2,\
3.25)$. Larger $\nu$ values have one stationary solution for all
$m\geq 2$ (see Appendix B), namely the $f_1$ root of expression
(\ref{roots}).
If $\nu$ is increased further, then the second root $f_2$ of
expression (\ref{roots}) also fulfills all necessary physical
requirements on $f$ parameter. Here, $f_2$ root is always smaller
than $f_1$ root and we expect that whenever there are two
theoretical solutions possible, then $f_2$ tends to be the more
realistic root because of its relatively lower values. The
limiting case of $\nu\to\infty$ represents tightly wound spiral
arms and the well-known WKBJ approximation (Lin \& Shu 1964, 1966)
should become applicable.

The curve $f_2(\lambda)$ shows that there is a one-to-one correspondence
between $f_2$ root and strong magnetic fields. Therefore, by
determining $f_2$ ratio through observations (e.g. rotation curves
or gravitational lensing effects) and equation (\ref{eq:1.a}), the
magnetic field strength can also be determined. For weak magnetic
fields, root $f_2$ does not vary significantly enough and thus,
isopedic magnetic field strength $B_z$ cannot be sensibly
determined as a result of uncertainties in galactic observations.

\subsection{Two Limiting Parameter Regimes}

The two limiting cases of $\lambda\to\infty$ and
$\delta_0\to\infty$ have been explored in the previous section. In
the limit of $\lambda\to\infty$, magnetic field almost vanishes
and one comes back to two gravitationally coupled hydrodynamic
disc system with an axisymmetric dark matter halo. For late-type
spiral galaxies, only perturbation configurations of $m=1$ lead to
stationary aligned solutions with a positive $f$ ratio. For
unaligned logarithmic perturbation configurations with $m\geq 2$,
the existence of stationary solutions also depends upon the choice
of dimensionless `radial wavenumber' $\nu$. For early-type spiral
galaxies of higher $\delta_0$, global stationary solutions of all
$m$ for perturbation configurations can be found. As already
noted, this implies the possibility of more numerous
multiple-armed spiral galaxies in the early Universe.

For $\delta_0\to\infty$ (see Fig. 5 for large $\delta_0$),
stationary perturbation solutions for $m\geq 2$ can be found for
every chosen parameter set. For $m=1$, solutions exist in a
certain range of $\lambda$. For example for $\nu=100$ and
$\delta_0=1000$, the case of $m=1$ has $f$ roots for
$\lambda=(1.2,\ 3.25)$ (see the right panel of Fig. 5).

To conclude for $\delta_0\to\infty$, for small $\nu$ or $\nu=0$,
all $m$ values have one $f$ root. But for larger $\nu$, the case
of $m\geq 2$ can have two $f$ roots. This is the same as that of
$\delta_0=0.05$. The difference lies in the values of $f$. The
larger the `radial wavenumber' $\nu$ value is, the larger are the
potential ratios $f_1$ and $f_2$.

For late-type spiral galaxies and unaligned logarithmic cases, a
weak magnetic field does not play a significant role as expected.
The younger the spiral galaxy is, the more important the role of a
magnetic field especially in the regime of stronger magnetic
fields. For very strong magnetic fields, it appears that $f$ ratio
is more closely connected to the magnetic field and the effect of
magnetic fields cannot be ignored. But for weak magnetic fields,
$f$ ratio remains more or less constant for all types of spiral
galaxies.

Our results can be summarized as follows. First of all, we
emphasize that the lopsided global configuration $m=1$ is an
exceptional case for which the curve $f(\lambda)$ has a distinctly
different shape as compared to those for the cases $m\geq 2$ (see
Appendix B). For both aligned and unaligned logarithmic spiral
cases with $m\geq 2$, strong magnetic fields (still realistic in
spiral galaxies) can bear a significant relation to the dark
matter halo in maintaining globally stationary perturbation
configurations.

Physically, we conclude that in spiral galaxies with or without
radial variations in perturbations, weak magnetic fields do not
influence stationary perturbation configurations. But when the
magnetic field strength is increased, a globally stationary
perturbation configuration would then become non-stationary if the
$f$ ratio is not high enough for the increased magnetic field
strength. A change of perturbation patterns might be possible
during the course of the galaxy evolution. For example, flocculent
galaxies might represent the transitional phase for global pattern
changes of galaxies. This prediction may be tested by numerical
simulations and by deep survey of morphological observations for
galaxies in the early Universe.


Our model analysis has shown that the magnetic field needs to be
sensibly chosen with other given disc parameters in order to
maintain a global perturbation configuration with a stationary
pattern in our frame of reference. Therefore, a variation or an
adjustment of $B_z$ may put a stationary perturbation
configuration into a non-stationary one or a non-stationary
perturbation configuration into a stationary one
because
there exists a one-to-one correspondence between the potential
ratio $f$ and a sufficiently strong magnetic field $B_z$ for
globally stationary perturbation configurations. For such a
stationary disc balance in general, the potential ratio $f$ can
vary in a considerably large range depending on the choice of
other relevant parameters in sensible regimes. As expected on
intuitive ground, sufficiently weak magnetic fields exert fairly
small influence on stationary perturbation configurations in our
composite model. The main reason is that the magnetic pressure and
tension together are not strong enough as compared with other
forces which play important dynamic roles in our composite
system.

In the regime of weak magnetic fields, there also exist
instabilities widely explored in the literature.
The magneto-Jeans instability (MJI) is an instability which is
based on background in-plane magnetic fields.
In recent years, several authors (e.g. Kim \& Ostriker 2001; Kim,
Ostriker \& Stone 2002; Shetty \& Ostriker 2006) have performed
numerical MHD simulations for galaxies. In particular, they
studied azimuthal magnetic fields that lie in the disc plane (see
also Lou \& Zou 2004, 2006; Lou \& Bai 2006). For our study of
isopedic magnetic fields, as already mentioned, Lou \& Wu (2005)
have proved that a constant $\lambda$ is a consequence of the
frozen-in condition on magnetic flux (see also Wu \& Lou 2006).
Our composite MHD model offers a two-dimensional description of an
isopedic magnetic field. Due to the idealization of razor-thin
discs, perturbations lie in the galactic plane and have no $\hat
z-$components. Hence, a treatment of the magnetorotational
instability (MRI) is impossible due to its necessary requirement
of a perturbation $e^{i k_z z}$ along $\hat z-$direction
(Chandrasekhar 1960; Balbus \& Hawley 1991, 1998; Balbus 2003).
When switching over to discs of finite thicknesses, the MRI must
be included and a larger $f$ is probably needed in order to
counteract MRIs. With our theory, the range of strong vertical
magnetic fields is well studied, whereas the effect of weak
magnetic fields is underestimated since we do not account for MRIs
which play an important role for weak magnetic fields. The results
that we obtain here cannot be directly applied to coplanar
magnetic fields, thus we can make no predictions for the MJIs.

The galactic application of our disc model results leads to two
methods which may be utilised to determine either the isopedic
magnetic field strength $B_z$ or the mass of an axisymmetric dark
matter halo. First, by using equation (\ref{eq:1.a}) through
observations of a disc galaxy, one can estimate the gravitational
potential ratio $f$. In the case of a stationary pattern of a
perturbation configuration, only one distribution of magnetic
field strengths is possible for this $f$ value. This is a new
approach of determining the distribution of isopedic magnetic
field strengths in disc galaxies. Proceeding in the opposite
direction, these model results may also be applied to determine
$f$ ratio by using observationally inferred distribution of
magnetic field strengths. This appears to be an alternative method
to estimate the halo mass of dark matter. The real challenge of
these procedures is to determine whether a perturbation pattern is
stationary or quasi-stationary through independent observational
diagnostics.
In practice, the relevant parameters that
need to be determined by galactic observations for our proposed
method to work are $m$, $\beta$ (the scaling index for radial
variations of unperturbed disc variables), $\beta_1^d$ (the
scaling index for radial variations of perturbation disc
variables), unperturbed disc surface mass densities
$\Sigma^{(i)}_0(r)$, effective sound speeds $a^{(i)}$, either
$\lambda$ (for the determination of gravitational potential ratio
$f$) or the other way around $f$ (for the determination of
distribution of magnetic field strengths $B_z$). It is indeed a
challenging task for observations to specifically identify some of
these parameters and as far as we know, galactic observations so
far still have not determined completely all these parameters with
error bars in one galaxy.

In addition, this work can also be understood as a preparation for
the following study on MHD density waves in such a composite disc
system since all relevant quantities and their relations among
each other are presented in this work. In reference to singular
isothermal disc (SID) models of Shu \& Li (1997), Shu et al.
(2000), Lou \& Shen (2003), Lou \& Zou (2004, 2006), Shen et al.
(2005), there exist two classes of solutions for stationary
magnetohydrodynamic (MHD) perturbation configurations with
in-phase and out-of-phase density perturbations in the two discs.
We expect for the case of a scale-free stellar disc and an
isopedically magnetized scale-free gaseous disc embedded in an
axisymmetric dark matter halo, that there are also two classes of
in-phase and out-of-phase density perturbations. The formulae and
results obtained here can be used later to analyze such MHD
density waves.

\section*{Acknowledgements}

This research has been supported in part by Deutscher Akademischer
Austauschdienst (DAAD; German Academic Exchange Service). This
research was supported in part
by the Tsinghua Centre for Astrophysics,
by NSFC grants 10373009 and 10533020 at Tsinghua University, and by
the SRFDP 20050003088 and 200800030071 and the Yangtze Endowment
from the Ministry of Education at Tsinghua University.

\appendix
\section{Derivations of stationary dispersion
relation and relevant equations}\label{sec:dr}

For coplanar MHD perturbations in a background rotating
axisymmetric disc system with
$v^{(i)}_r=v^{(i)}_{r0}+v^{(i)}_{r1}$ , $\
v^{(i)}_\theta=v^{(i)}_{\theta 0}+ v^{(i)}_{\theta 1}$ , $\
\Sigma^{(i)}=\Sigma^{(i)}_0+\Sigma^{(i)}_1$ , $\
\Phi^{(i)}=\Phi^{(i)}_0+\Phi^{(i)}_1$ and $\
H^{(i)}=H^{(i)}_0+H^{(i)}_1$ in equations
$(\ref{eq:conti})-(\ref{eq:euler2})$, where superscript $i$ within
parentheses can be set as $s$ and $g$ for stellar and magnetized
gaseous discs respectively, it is straightforward to obtain
linearized perturbation equations below
\begin{eqnarray}
\frac{\partial\Sigma^{(i)}_1}{\partial t}+\frac{1}{r}
\frac{\partial}{\partial r}\left[r\Sigma^{(i)}_0v^{(i)}_{r
1}\right]+
\frac{1}{r}\frac{\partial}{\partial\theta}\left[\Sigma^{(i)}_0
v^{(i)}_{\theta 1}+\Sigma^{(i)}_1v^{(i)}_{\theta 0}\right]=
0 \label{eq:1. Gleichung}\ ,\\
\frac{\partial v^{(s)}_{r 1}}{\partial t}+\Omega^{(s)}_0
\frac{\partial v^{(s)}_{r 1}}{\partial \theta} -2 \Omega^{(s)}_0
v^{(s)}_{\theta 1}=-\frac{\partial\left[H^{(s)}_1+\Phi^{(s)}_1
+\Phi^{(g)}_1\right]}{\partial r}\ ,\label{A2}\\
\frac{\partial v^{(g)}_{r 1}}{\partial t}+\Omega^{(g)}_0
\frac{\partial v^{(g)}_{r 1}}{\partial\theta}-2\Omega^{(g)}_0
v^{(g)}_{\theta 1}=-\frac{\partial\left[\Theta
H^{(g)}_1+\Phi^{(s)}_1
+\epsilon\Phi^{(g)}_1\right]}{\partial r}\ ,\label{A3}\\
\frac{\partial v^{(s)}_{\theta 1}}{\partial t}+v^{(s)}_{r 1}
\bigg[\Omega^{(s)}_0 +r\frac{\rmn{d}\Omega^{(s)}_0}{\rmn{d}
r}\bigg]+\Omega^{(s)}_0\frac{\partial v^{(s)}_{\theta 1}}{\partial
\theta}+ \Omega^{(s)}_0 v^{(s)}_{r 1}=-\frac{\partial
\left[H^{(s)}_1+\Phi^{(s)}_1
+\Phi^{(g)}_1\right]}{r\partial \theta}\ ,\label{A4}\\
\frac{\partial v^{(g)}_{\theta 1}}{\partial t} +v^{(g)}_{r 1}
\bigg[\Omega^{(g)}_0+r\frac{\rmn{d} \Omega^{(g)}_0}{\rmn{d}
r}\bigg]+\Omega^{(g)}_0\frac{\partial v^{(g)}_{\theta 1}}{\partial
\theta}+\Omega^{(g)}_0 v^{(g)}_{r 1}=-\frac{\partial\left[\Theta
H^{(g)}_1+\Phi^{(s)}_1+\epsilon
\Phi^{(g)}_1\right]}{r\partial\theta}\ ,\label{eq:letzte Gleichung}
\end{eqnarray}
where $v^{(i)}_{r0}=0$ for the background disc configuration.
Equation (\ref{eq:1. Gleichung}) represents the two mass
conservation equations, equations (\ref{A2}) and (\ref{A3}) are
the two radial momentum equations, and equations (\ref{A4}) and
(\ref{eq:letzte Gleichung}) are the two azimuthal momentum
equations; effects of the isopedic magnetic field are subsumed in
the two parameters $\Theta$ and $\epsilon$ (see equations
\ref{epsilon} and \ref{Btheta}) for the magnetized gaseous disc in
momentum equations (\ref{A3}) and (\ref{eq:letzte Gleichung}); and
the gravitational coupling between the stellar and gaseous discs
is represented by the sum of gravitational potential perturbations
$\Phi^{(s)}_1+\Phi^{(g)}_1$ or $\Phi^{(s)}_1+\epsilon\Phi^{(g)}_1$
in two-dimensional perturbation equations (\ref{A2}) to
(\ref{eq:letzte Gleichung}). As the axisymmetric gravitational
potential of a massive dark matter halo is presumed to be
unperturbed for simplicity, its perturbation does not appear on
the right-hand sides of equations (\ref{A2}) to (\ref{eq:letzte
Gleichung}).\footnote{The assumption of an unperturbed
axisymmetric dark matter halo corresponds to a limiting case. It
would be interesting to explore the case where the dark matter
halo is perturbed and coupled to disc perturbations
gravitationally.} For perturbations cast in the form of Fourier
decomposition, we write for example
\begin{eqnarray}
v^{(i)}_{r 1}=A^{(i)}(r)
\exp [i(\omega t-m \theta)]\ , \label{eq:v_rst"orung}\\
v^{(i)}_{\theta 1}=B^{(i)}(r)\exp [i(\omega t-m \theta)]\ ,
\label{eq:v_thetast"orung}
\end{eqnarray}
where $m$ is an integral number to characterize azimuthal
variations, $\omega$ is the angular perturbation frequency,
$A^{(i)}(r)$ and $B^{(i)}(r)$ are small-magnitude functions of
radius $r$, and superscripts $i=s,\ g$ stand for associations with
stellar and gaseous discs, respectively. Two new frequency
parameters are now introduced below for simplicity. The first one
is the abbreviation
$\overline{\omega}^{(i)}\equiv\omega-m\Omega^{(i)}_0$ and the
second one is the so-called epicyclic frequency $\kappa^{(i)}_0$
of a rotating disc defined by
\begin{equation}
\left[\kappa^{(i)}_0\right]^2=r\rmn{d}\left[\Omega^{(i)}_0\right]^2
/\rmn{d} r+4 \left[\Omega^{(i)}_0\right]^2
=2\left[b^{(i)}_0\right]^2(1-\beta)
r^{-2\beta-2}=2(1-\beta)\left[\Omega^{(i)}_0\right]^2\ .
\end{equation}
Substituting all expressions of Fourier harmonics into equations
(\ref{eq:1. Gleichung})$-$(\ref{eq:letzte Gleichung}), we readily
obtain the following perturbation equations
\begin{eqnarray}
\Sigma^{(i)}_1i\overline{\omega}^{(i)}+\frac{1}{r}
\frac{\partial}{\partial r}\left[r\Sigma^{(i)}_0v^{(i)}_{r
1}\right]- \frac{i m}{r}\Sigma^{(i)}_0v^{(i)}_{\theta 1}=0\ ,
\label{eq:continuity} \\
v^{(s)}_{r 1}i\overline{\omega}^{(s)}-2\Omega^{(s)}_0
v^{(s)}_{\theta 1}=-\frac{\partial\left[H^{(s)}_1+\Phi^{(s)}_1
+\Phi^{(g)}_1\right]}{\partial r}\ ,\label{eq:v_r1}\\
v^{(g)}_{r,1} i\overline{\omega}^{(g)}-2\Omega^{(g)}_0
v^{(g)}_{\theta 1}=-\frac{\partial\left[\Theta
H^{(g)}_1+\Phi^{(s)}_1
+\epsilon \Phi^{(g)}_1\right]}{\partial r}\ ,\\
v^{(s)}_{\theta 1} i\overline{\omega}^{(s)}+v^{(s)}_{r 1}
\frac{\left[\kappa^{(s)}_0\right]^2}{2\Omega^{(s)}_0}=
\frac{i m \left[H^{(s)}_1+\Phi^{(s)}_1+\Phi^{(g)}_{1}\right]}{r}\ ,\\
v^{(g)}_{\theta 1}i\overline{\omega}^{(g)}+v^{(g)}_{r 1}
\frac{\left[\kappa^{(g)}_0\right]^2}{2\Omega^{(g)}_0}=\frac{i m
\left[\Theta
H^{(g)}_1+\Phi^{(s)}_1+\epsilon\Phi^{(g)}_{1}\right]}{r}\ ,
\label{eq:v_theta2}
\end{eqnarray}
for both stellar and isopedically magnetized gaseous discs coupled
by gravity. Furthermore, the perturbation of surface mass density is
set to the following form of Fourier harmonics
\begin{equation}
\Sigma^{(i)}_1= S^{(i)}_1r^{-2\beta_1-1}\exp[i(\omega t-m\theta
+\nu\ln r)]=S^{(i)}_1r^{-2\beta^d_1-1}\exp[i(\omega t-m\theta)]\
\label{eq:Sigma_1}
\end{equation}
with $S^{(i)}_1 $ being small-magnitude coefficients to justify the
perturbation approach and
\begin{equation}
\beta^d_1\equiv\beta_1-i\nu/2\ \label{eq:beta_1^d}
\end{equation}
being complex in general for $\nu\neq 0$.
The unit complex factor $\exp (-im\theta)$ represents periodic
azimuthal variations and the unit complex factor $\exp (i\nu\ln
r)$ represents radial variations with $\nu$ closely related to the
radial wavenumber. In general, $\beta_1$ parameter for
perturbations can be different from $\beta$ parameter used for
characterizing the background axisymmetric equilibrium disc in
rotation. According to several formulae derived by Qian (1992),
the corresponding perturbation of gravitational potential
associated with a surface mass density perturbation is simply
given by
\begin{equation}
\Phi_1^{(i)}=-GY_m(\beta^d_1)S^{(i)}_1r^{-2\beta^d_1}\exp[i(\omega
t-m\theta)] =-GrY_m(\beta^d_1)\Sigma^{(i)}_1
\end{equation}
analytically and the corresponding perturbation of enthalpy is given
by
\begin{equation}
H^{(i)}_1= \frac{\rmn{d} H^{(i)}_0}{\rmn{d}
\Sigma^{(i)}_0}\Sigma^{(i)}_1=n
k^{(i)}\left[\Sigma^{(i)}_0\right]^{n-2} \Sigma^{(i)}_1
=\frac{\left[a^{(i)}\right]^2}{\Sigma^{(i)}_0}\Sigma^{(i)}_1\ .
\end{equation}
It follows immediately that
\begin{eqnarray}
H^{(s)}_1+\Phi^{(s)}_1+\Phi^{(g)}_1=Gr\Sigma^{(s)}_1 K^{(s)}\ ,
\hspace{1 cm}\rmn{with}\hspace{1 cm}K^{(s)}\equiv
\frac{\left[a^{(s)}\right]^2}{G r\Sigma^{(s)}_0}
-Y_m(\beta_1^d)(1+\delta_1)\ ;\label{Ks}\\
\Theta H^{(g)}_1+\Phi^{(s)}_1+\epsilon\Phi^{(g)}_1=G r
\Sigma^{(g)}_1 K^{(g)}\ ,\hspace{1 cm} \rmn{with} \hspace{1 cm}
K^{(g)}\equiv\frac{\Theta\left[a^{(g)}\right]^2}{G r
\Sigma^{(g)}_0}-Y_m(\beta_1^d)(\epsilon+\delta_1^{-1})\ ,\label{Kg}
\end{eqnarray}
where $\delta_1\equiv\Sigma_1^{(g)}/\Sigma_1^{(s)}$ is a constant
ratio for surface mass density perturbations by expression
(\ref{eq:Sigma_1}), and $K^{(s)}$ and $K^{(g)}$ defined by
equations (\ref{Ks}) and (\ref{Kg}) respectively are two
dimensionless constants because $\big[a^{(i)}\big]^2\propto
r^{-2\beta}$ and $r \Sigma^{(i)}_0\propto r^{-2\beta}$. For
coplanar perturbations in the stellar disc in rotation, we then
have the following perturbation equations
\begin{eqnarray}
\Sigma^{(s)}_1 i\overline{\omega}^{(s)}+\frac{1}{r}
\frac{\partial}{\partial r}\left[r\Sigma^{(s)}_0v^{(s)}_{r
1}\right]- \frac{i m}{r} \Sigma^{(s)}_0 v^{(s)}_{\theta 1}=0\ ,
\label{eq:1stellar} \\
v^{(s)}_{r 1} i\overline{\omega}^{(s)}-2\Omega^{(s)}_0
v^{(s)}_{\theta 1}=-GK^{(s)}\frac{\partial}{\partial r}
\left[\Sigma^{(s)}_1 r\right]\ ,\label{eq:2stellar}\\
v^{(s)}_{\theta 1} i\overline{\omega}^{(s)}+v^{(s)}_{r 1}
\frac{\left[\kappa^{(s)}_0\right]^2}{2\Omega^{(s)}_0}= i m G
\Sigma^{(s)}_1 K^{(s)}\ .\label{eq:3stellar}
\end{eqnarray}
Meanwhile for coplanar MHD perturbations in the isopedically
magnetized gaseous disc in rotation, we obtain in parallel
\begin{eqnarray}
\Sigma^{(g)}_1 i\overline{\omega}^{(g)}+\frac{1}{r}
\frac{\partial}{\partial r}\left[r\Sigma^{(g)}_0 v^{(g)}_{r
1}\right]-\frac{i m}{r}\Sigma^{(g)}_0v^{(g)}_{\theta 1}=0\ ,
\label{eq:1gaseous}\\
v^{(g)}_{r 1} i\overline{\omega}^{(g)}-2\Omega^{(g)}_0
v^{(g)}_{\theta 1}=-GK^{(g)}\frac{\partial}{\partial r}
\left[\Sigma^{(g)}_1 r\right]\ ,\label{eq:2gaseous}\\
v^{(g)}_{\theta 1} i\overline{\omega}^{(g)}+v^{(g)}_{r 1}
\frac{\left[\kappa^{(g)}_0\right]^2}{2\Omega^{(g)}_0}=i mG
\Sigma^{(g)}_1 K^{(g)}\ .\label{eq:3gaseous}
\end{eqnarray}
>From equations $(\ref{eq:2stellar})-(\ref{eq:3stellar})$ and
$(\ref{eq:2gaseous})-(\ref{eq:3gaseous})$, we directly derive
expressions for radial and azimuthal velocity perturbations
$v^{(i)}_{r 1}$ and $v^{(i)}_{\theta 1}$ as given below.
\begin{eqnarray}
v^{(i)}_{r 1}=\frac{iGK^{(i)}}
{\big[\overline{\omega}^{(i)}\big]^2-\big[\kappa_0^{(i)}\big]^2}
\left(-\frac{2m\Omega_0^{(i)}}{r}+\overline{\omega}^{(i)}
\frac{\partial}{\partial r}\right)
\left[\Sigma^{(i)}_1 r\right]\ ,\\
v^{(i)}_{\theta 1}=\frac{GK^{(i)}}
{\big[\overline{\omega}^{(i)}\big]^2-\big[\kappa_0^{(i)}\big]^2}
\left( \frac{m\overline{\omega}^{(i)}}{r}-
\frac{\big[\kappa_0^{(i)}\big]^2}{2
\Omega_0^{(i)}}\frac{\partial}{\partial r}\right)
\left[\Sigma^{(i)}_1r\right]\ .
\end{eqnarray}
Substituting these expressions into the mass conservation
equations and using expression (\ref{eq:Sigma_1}), we finally
arrive at
\begin{equation}
\frac{\overline{\omega}^{(i)}r}{GK^{(i)}
\Sigma^{(i)}_0}=\frac{\big[m^2-4(\beta^d_1)^2+2\beta^d_1\big]
\overline{\omega}^{(i)}-2m\beta
\Omega_0^{(i)}}{\big[\overline{\omega}^{(i)}\big]^2-\big[\kappa_0^{(i)}\big]^2
}- \frac{4(\beta+1)\big[m\Omega_0^{(i)}+\beta^d_1
\overline{\omega}^{(i)}\big]\overline{\omega}^{(i)}\omega} {\left[
[\overline{\omega}^{(i)}]^2-[\kappa_0^{(i)}]^2\right]^2}\equiv
Q^{(i)}\ \label{eq:drdiscs}
\end{equation}
for both stellar and magnetized gaseous discs, respectively, and
for defining the corresponding $Q^{(i)}$. A combination of these
two equations for perturbations in stellar and gaseous discs leads
to the dispersion relation of the gravity coupled disc
configuration, namely
\begin{equation}
\left\{\left[a^{(s)}\right]^2-G\Sigma^{(s)}_0 rY_m(\beta^d_1)
-\overline{\omega}^{(s)}r^2/Q^{(s)}\right\}\left\{\Theta
\left[a^{(g)}\right]^2-\epsilon G\Sigma^{(g)}_0r
Y_m(\beta^d_1)-\overline{\omega}^{(g)}r^2/Q^{(g)}\right\}=
G\Sigma^{(s)}_0 rY_m(\beta^d_1)G\Sigma^{(g)}_0 r Y_m(\beta^d_1)\
.\label{eq:dispersionrelation}
\end{equation}
Relation (\ref{eq:dispersionrelation}) is comprehensive and it
contains all useful information about the disc dynamics after a
perturbation has arisen there. Depending on the purpose of
investigation, one can define different parameter regimes by using
this dispersion relation. We study non-axisymmetric stationary
perturbations with $m\neq 0$ and $\omega=0$ which can be either
aligned configurations or unaligned logarithmic spirals. For the
stationary case of $\omega=0$, equation
(\ref{eq:dispersionrelation}) reduces to
\begin{eqnarray}
\left\{\big[a^{(s)}\big]^2-G\Sigma^{(s)}_0 r Y_m(\beta^d_1) -
\big[\Omega_0^{(s)}\big]^2r^2\left[\frac{m^2-2(1-\beta )}{m^2-4
(\beta^d_1 )^2+2\beta^d_1
+2\beta} \right]\right\}
\qquad\qquad\qquad\qquad\qquad\qquad\qquad\qquad\qquad\qquad \nonumber \\
 \qquad\times\left\{\Theta \big[a^{(g)}\big]^2-\epsilon G \Sigma^{(g)}_0 r
Y_m(\beta^d_1)-\big[\Omega_0^{(g)}\big]^2 r^2 \left[
\frac{m^2-2(1-\beta )}{m^2-4 (\beta^d_1)^2+2 \beta^d_1 +2\beta }
\right] \right\} = \left[G\Sigma^{(s)}_0 r
Y_m(\beta^d_1)\right]\left[G \Sigma^{(g)}_0 r Y_m(\beta^d_1)\right]\
,\label{eq:drw=0}
\end{eqnarray}
where, according to expressions (\ref{eq:1.a}) and (\ref{eq:2.a})
for the background equilibrium disc system, we have disc angular
rotation speeds given by
\begin{equation}
\Omega_0^{(s)}\equiv v^{(s)}_{\theta 0}/r\ \hspace {0.5cm}
\rmn{with}\hspace {0.5cm}\left[v^{(s)}_{\theta 0}\right]^2=2\beta r
GY_0(\beta)\left[\Sigma^{(s)}_0+\Sigma^{(g)}_0\right](1+f)-
\left[a^{(s)}\right]^2(2\beta+1)\ ,\label{eq:omegastellar}
\end{equation}
\begin{equation}
\Omega^{(g)}_0\equiv v^{(g)}_{\theta 0}/r\ \hspace {0.5cm}
\rmn{with}\hspace {0.5cm}\left[v^{(g)}_{\theta 0}\right]^2 =2\beta r
GY_0(\beta)
\left[\big[\Sigma^{(s)}_0+\Sigma^{(g)}_0\big](1+f)-(1-\epsilon)
\Sigma^{(g)}_0\right]-\Theta\left[a^{(g)}\right]^2(2\beta+1)\
.\label{eq:omegagaseous}
\end{equation}
As the gravitational potential ratio $f$ should be the same in
both equations (\ref{eq:1.a}) and (\ref{eq:2.a}), a relation
between $\big[v^{(s)}_{\theta 0}\big]^2$ and $\big[v^{(g)}_{\theta
0}\big]^2$ can therefore be established, namely
\begin{equation}
\big[v^{(s)}_{\theta 0}\big]^2= \big[v^{(g)}_{\theta
0}\big]^2+(2\beta+1)\big\{\Theta\big[a^{(g)}\big]^2
-\big[a^{(s)}\big]^2\big\}+2\beta rGY_0(\beta)(1-\epsilon)
\Sigma^{(g)}_0\ .\label{eq:v^{(s)}v^{(g)}}
\end{equation}
Using expressions $v^{(i)}_{\theta 0}=b^{(i)}_0r^{-\beta}$
and $\Sigma^{(i)}_0=S^{(i)}_0 r^{-2\beta-1}$ for the background
scale-free disc system, equation (\ref{eq:v^{(s)}v^{(g)}}) can be
written as
\begin{equation}
\big[b^{(s)}_{0}\big]^2=\big[b^{(g)}_{0}\big]^2+(2\beta+1)\big\{\Theta
\big[a^{(g)}\big]^2 -\big[a^{(s)}\big]^2\big\}r^{2 \beta}+2 \beta
G Y_0(\beta) (1-\epsilon) S^{(g)}_0\ \label{eq:b^{(s)}(b^{(g)})}
\end{equation}
without radial dependence. We now introduce the rotational Mach
number $D^{(s)}$ for the stellar disc and the rotational
magnetosonic Mach number $D^{(g)}$ for the isopedically magnetized
gaseous disc, respectively.
\begin{eqnarray}
D^{(s)}\equiv\frac{v^{(s)}_{\theta 0}}{a^{(s)}}
=\frac{b^{(s)}_0}{a^{(s)} r^{\beta}}\ , \hspace{1 cm} \rightarrow
\hspace{1 cm} b^{(s)}_0=a^{(s)} r^{\beta}D^{(s)}\ ;\\
D^{(g)}\equiv\frac{v^{(g)}_{\theta 0}}{\sqrt{\Theta} a^{(g)}}
=\frac{b^{(g)}_0}{\sqrt{\Theta}a^{(g)}r^{\beta}}\ ,\hspace{1 cm}
\rightarrow \hspace{1 cm} b^{(g)}_0=\sqrt{\Theta}a^{(g)} r^{\beta}
D^{(g)}\ .
\end{eqnarray}
Using background disc equilibrium equations (\ref{eq:1.a}),
(\ref{eq:2.a}) and (\ref{eq:b^{(s)}(b^{(g)})}), the following
relations can then be derived,
\begin{eqnarray}
\big[D^{(s)}\big]^2=\frac{2\beta G Y_0(\beta) S^{(s)}_0
(1+\delta_0)(1+f)}
{\big[a^{(s)}\big]^2 r^{2\beta}}-(2\beta+1)\ ,\label{eq:D^{(s)}} \\
\big[D^{(g)}\big]^2=\frac{2\beta G Y_0(\beta)S^{(g)}_0 \big[ f
(1+\delta_0^{-1})+\delta^{-1}_0+\epsilon\big]
}{\Theta \big[a^{(g)}\big]^2 r^{2\beta}}-(2\beta+1)\ ,\label{eq:D^{(g)}}\\
\big[D^{(s)}\big]^2=\big[D^{(g)}\big]^2\Delta+(2\beta+1)(\Delta
-1)+2\beta GY_0(\beta)(1-\epsilon)S^{(g)}_0\big/\big[[a^{(s)}]^2
r^{2\beta}\big]\ ,\label{eq:connection D^{(s)},D^{(g)}}
\end{eqnarray}
with ratio $\Delta\equiv\Theta
\big[a^{(g)}\big]^2\big/\big[a^{(s)}\big]^2$. With these defined
quantities and relations (\ref{eq:D^{(s)}}), (\ref{eq:D^{(g)}})
and (\ref{eq:connection D^{(s)},D^{(g)}}), stationary dispersion
relation (\ref{eq:drw=0}) can be cast into
\begin{eqnarray}
\left\{1-\frac{GS^{(s)}_0Y_m(\beta^d_1)}{\big[a^{(s)}\big]^2
r^{2\beta}}-\left[D^{(s)}\right]^2\left[\frac{m^2-2(1-\beta )}{m^2-
4(\beta^d_1)^2 +2 \beta^d_1+2\beta}\right]\right\}
\qquad\qquad\qquad\qquad\qquad\qquad\qquad\qquad\nonumber \\
\times\left\{1-\frac{\epsilon G S^{(g)}_0 Y_m(\beta^d_1)}{\Theta
\big[a^{(g)}\big]^2r^{2\beta}}-
\left[D^{(g)}\right]^2\left[\frac{m^2-2(1-\beta
)}{m^2-4(\beta^d_1)^2 +2 \beta^d_1+2\beta
}\right]\right\}=\frac{GS^{(s)}_0 Y_m(\beta^d_1)}
{\big[a^{(s)}\big]^2r^{2\beta}}\frac{G
S^{(g)}_0Y_m(\beta^d_1)}{\Theta \big[a^{(g)}\big]^2r^{2\beta}}\ .
\end{eqnarray}

\subsection{Two Cases of Real Stationary Dispersion Relation
and Relevant Mathematical Expressions}
\label{sec:f,lambda}

Our main purpose is to establish relations between dark matter
halo (represented by $f$ ratio) and an isopedic magnetic field in
gaseous disc (represented by the isopedic ratio $\lambda$). For
the two cases of $\beta_1^d=\beta_1=\beta$ (with $\nu=0$) and of
$\beta_1=1/4$ (with $\nu\neq 0$), we show in the following that
$M$ and $Y_m$ as defined by equations (\ref{eq:M}) and (\ref{Ym})
become real numbers. In the literature, $\beta_1^d=\beta$ is
called the aligned case since the radial wavenumber parameter
$\nu$ vanishes in this case and thus, only azimuthal and no radial
wave variations are present. In this case, we have
\begin{eqnarray}
M=\frac{m^2-2(1-\beta )}{m^2+4\beta (1-\beta)}\ ,
\qquad\qquad\qquad\qquad\quad\ \ \\
Y_m(\beta)=\frac{\pi\Gamma(m/2-\beta+1/2)
\Gamma(m/2+\beta)}{\Gamma(m/2-\beta+1)\Gamma(m/2+\beta+1/2)}\ .
\end{eqnarray}
For $\nu\neq 0$ and thus a complex $\beta_1^d$, the general
expressions for $M$ and $Y_m(\beta_1^d)$ are
\begin{eqnarray}
M=\frac{m^2-2(1-\beta)}{m^2+2(\beta+\beta_1
-2\beta_1^2+\nu^2/2)+i\nu(4\beta_1-1)}\ ,\qquad\qquad\quad\label{Mnu}\\
Y_m(\beta_1^d)=\frac{\pi\Gamma(m/2-\beta_1+i\nu/2+1/2)
\Gamma(m/2+\beta_1-i\nu/2)}{\Gamma(m/2-\beta_1+i\nu/2+1)
\Gamma(m/2+\beta_1-i\nu/2+1/2)}\ .\label{Ymnu}
\end{eqnarray}
By setting $\beta_1=1/4$ in expressions (\ref{Mnu}) and
(\ref{Ymnu}), we immediately have
\begin{eqnarray}
M=\frac{m^2-2(1-\beta)}{m^2+2\beta+1/4+\nu^2}\ ,\qquad\qquad\\
Y_m(\beta_1^d)=\left
|\frac{\Gamma(m/2+1/4+i\nu/2)}{\Gamma(m/2+3/4+i\nu/2)}\right |^2\
.\label{Ybt1}
\end{eqnarray}
Expression (\ref{Ybt1}) is based on the relation
$\Gamma(z)^*=\Gamma(z^*)$ for Gamma-function where $^{*}$ denotes
the complex conjugate operation. For these two cases
$\beta_1^d=\beta_1=\beta$ (with $\nu=0$) and of $\beta_1=1/4$
(with $\nu\neq 0$ in general), we explore numerically the
behaviour $f(\lambda)$ by specifying other relevant model
parameters.


\section{Dependence of gravity potential ratio $f$
on perturbation orders $m$}\label{sec:diskrepancy}

Here, we briefly describe the gravitational potential ratio $f$ as
a function for different $m$ values and provide an upper limit for
the difference of $f(m+2)-f(m)$. In general, the function $f$
depends on $m$ through functional parameters $M$ and
$Y_m(\beta_1^d)$.
In the limit of $m\to\infty$, we have $M\to 1$ and therefore the
difference $\Delta M=M(m+1)-M(m)\to 0$. For the property of
functional parameter $Y_m(\beta_1^d)$, we first discuss the
property of Gamma-function. For Gamma functions, the well-known
recursion formula is simply
\begin{equation}
\Gamma(x+1)=x\Gamma(x)\ ,
\end{equation}
which immediately leads to the following four relations
\begin{eqnarray}
\Gamma ((m+2)/2-\beta_1^d+1/2)=(m/2-\beta_1^d+1/2)
\Gamma (m/2-\beta_1^d+1/2)\ ,\\
\Gamma ((m+2)/2+\beta_1^d) = (m/2+\beta_1^d)
\Gamma(m/2+\beta_1^d)\ , \\
\Gamma ((m+2)/2-\beta_1^d+1) = (m/2-\beta_1^d+1)
\Gamma(m/2-\beta_1^d+1)\ , \\
\Gamma ((m+2)/2+\beta_1^d+1/2) = (m/2+\beta_1^d+1/2)
\Gamma(m/2+\beta_1^d+1/2) \
\end{eqnarray}
as applied to our case under consideration. Therefore, we have
\begin{equation}
Y_{m+2}(\beta^d_1)=\frac{\big(m-2\beta_1^d+1\big)\big(m+2\beta_1^d\big)}{\big(m-2
\beta_1^d+2\big)\big(m+2\beta_1^d+1\big)}Y_{m}(\beta^d_1)\
.\label{eq:Y_m+2}
\end{equation}
For $m\to\infty$, it follows that $Y_{m+2}(\beta_1^d)\to
Y_m(\beta_1^d)$, indicating the difference $\Delta
Y_m(\beta_1^d)=Y_{m+2}(\beta^d_1)-Y_m(\beta_1^d)\to 0$.
Based on these analyses, we conclude that for very large values of
$m$, there is no significant difference between $f(m+2)$ and
$f(m)$, and therefore $f(m+2)\to f(m)$ in the limit of large $m$
values.

For $m=2$, we have from equations (\ref{eq:M}) and (\ref{eq:Y_m+2})
\begin{eqnarray}
M(4)-M(2)=\frac{7+\beta }{8-2(\beta^d_1 )^2
+\beta^d_1+\beta}-\frac{1+\beta }{2-2(\beta^d_1)^2
+\beta^d_1+\beta}\ ,\label{DM}\\
Y_4(\beta^d_1)-Y_2(\beta^d_1)=\left[\frac{(3-2\beta_1^d)(1+
\beta_1^d)}{(2-\beta_1^d)(3+2\beta_1^d)}-1\right]Y_2(\beta^d_1)=
\left[\frac{3+\beta_1^d-2(\beta_1^d)^2}{6+\beta_1^d-2
(\beta_1^d)^2}-1\right]Y_2(\beta^d_1)\equiv NY_2(\beta^d_1)\
,\label{Ndef}
\end{eqnarray}
\begin{figure}
\begin{center}
\includegraphics[angle=0, scale=0.95]{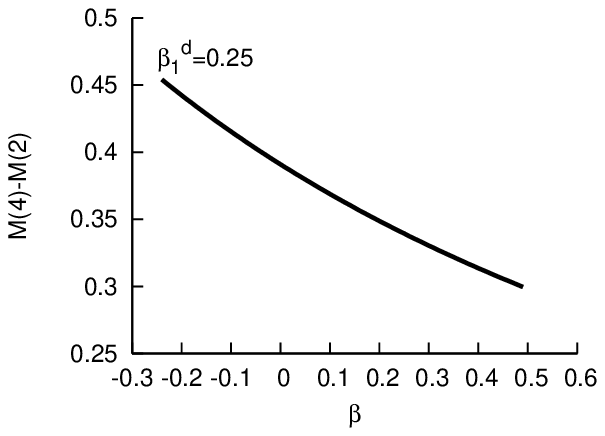}
\includegraphics[angle=0, scale=0.95]{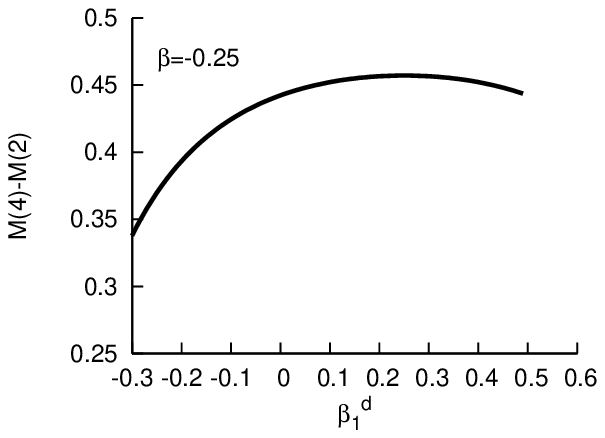}
\caption{For the left panel, the difference of $M(4)-M(2)$
is shown as a function of $\beta$ parameter with a fixed real
$\beta_1^d$, with a maximum at $\beta=-0.25$ and $\beta_1^d \sim
0.25$; for a real value of $\beta_1^d$, the difference $M(4)-M(2)$
appears to be a monotonically decreasing function with increasing
$\beta$ value.
For the right panel, the difference of $M(4)-M(2)$ is shown
as a function of real $\beta_1^d$ values with a fixed value
$\beta=-0.25$. A maximum value can be readily identified. }
\label{fig:M(beta)}
\end{center}
\end{figure}
which defines the coefficient $N$ as a function of $\beta^d_1$
that is complex for $\nu\neq 0$. The difference $M(4)-M(2)$ is a
function of both $\beta$ and $\beta^d_1$.
The largest difference is found for $\beta=-0.25$ and
$\beta_1^d\sim 0.2-0.3$ with $\Delta M\equiv M(4)-M(2)=0.46$ (see
Figure \ref{fig:M(beta)}). For applications to spiral galaxies,
only positive $\beta$ values are physically relevant. But for a
theoretical consideration, a negative $\beta$ is allowed to
estimate limit.
To conclude, we obtain
\begin{equation}
\Delta M\leq 0.46\ .
\end{equation}
\begin{figure}
\begin{center}
\includegraphics[angle=0, scale=0.95]{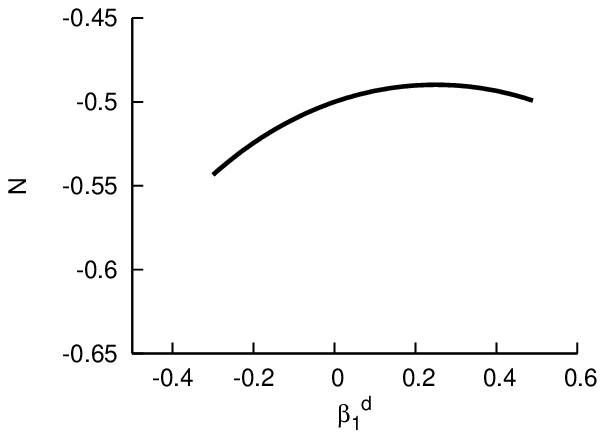}
\caption{Variation of coefficient $N(\beta_1^d)$ defined by
equation (\ref{Ndef}) as a function of real $\beta_1^d$ values.
The factor $N(\beta_1^d)$ in relation $Y_4=(N+1)Y_2$ as a function
of $\beta_1^d$. }\label{fig:N(beta)}
\end{center}
\end{figure}
For a real range of $\beta_1^d\in [-0.5, 0.5]$ with $\nu=0$,
functional parameter $N(\beta_1^d)$ is plotted in Figure
\ref{fig:N(beta)}. The smallest $N$ is $-0.6$ for $\beta_1^d
=-0.5$. From equation (\ref{eq:Y_m+2}), one can also see that
$Y_{m+2}(\beta^d_1)$ is always smaller than $Y_m(\beta^d_1)$. This
fact explains the negative values shown in Figure
\ref{fig:N(beta)}. For the absolute difference, we get inequality
\begin{equation}
|\Delta Y_m|=|Y_{m+2}-Y_m|\leq 0.6Y_m\ .
\end{equation}
In contrast to $\Delta M$ which can be determined as a constant,
$\Delta Y_m$ is a function of $Y_m$. For $\Delta f$, the upper
limit is $f(4)-f(2)$ which is different each time depending on the
chosen parameters. As a result, the difference of two functions
$f(n)$ and $f(n+2)$ with $n=m,m+1,\cdots$ goes towards zero,
namely
\begin{eqnarray}
f(m+2)-f(m)\to 0\hspace{1cm}\rmn{for}\hspace{1cm} m\to\infty\ .
\end{eqnarray}
\begin{eqnarray}
f(m+3)-f(m+1)\to 0\hspace{1cm}\rmn{for}\hspace{1cm} m\to\infty\ .
\end{eqnarray}

\section{Perturbation density ratio $\delta_1$ in the model disc system}

In order to compare the mass density perturbations in the two
discs, the ratio $\delta_1=\Sigma_1^{(g)}/\Sigma_1^{(s)}$ is
derived in the following by using equation (\ref{eq:drdiscs}).
This ratio $\delta_1$ is contained implicitly in $K^{(i)}$
expressions. Using this relation for the stellar disc and
$\omega^{(g)}=0$, we derive
\begin{equation}
\delta_1=\frac{\Sigma^{(g)}_1}{\Sigma^{(s)}_1}
=\frac{-\big[D^{(s)}\big]^2\big[A^{(s)}\big]^2
\big[m^2-2(1-\beta)\big]}{GS^{(s)}_0 Y_m(\beta^d_1)
\big[m^2-4(\beta^d_1)^2 +2\beta^d_1 +2\beta\big]}+
\frac{\big[A^{(s)}\big]^2}{G S^{(s)}_0 Y_m(\beta^d_1)}-1\
.\label{C1}
\end{equation}
We substitute expression (\ref{eq:connection D^{(s)},D^{(g)}}) of
$D^{(s)}$ into equation (\ref{C1}) to relate the two discs. With
$N=-(m^2-2+2\beta )\ ,$ we then obtain
\begin{equation}
\delta_1=N\frac{(D^{(g)})^2\Theta (A^{(g)})^2
+(2\beta+1)\big[\Theta (A^{(g)})^2-(A^{(s)})^2\big]+2\beta G
Y_0(\beta)(1-\epsilon)S^{(g)}_0}{GY_m(\beta_1^d)S^{(s)}_0
\big[m^2-4(\beta^d_1)^2+2\beta^d_1
+2\beta\big]}+\frac{(A^{(s)})^2} {GS^{(s)}_0Y_m(\beta_1^d)}-1\
.\label{eq:delta_1}
\end{equation}
The ratio $\delta_1=\Sigma^{(g)}_1/\Sigma^{(s)}_1$ is negative in
the case of out-of-phase density wave perturbations. For this
case, the gravity effect is weaker and the MHD density wave speed
is faster. In the case of in-phase density wave perturbations,
this ratio $\delta_1$ is positive, the gravity effect is stronger
and the MHD density wave speed is slower (Lou \& Fan 1998).
\label{lastpage}
\end{document}